%% file: main.tex
\documentclass[conference,compsoc]{IEEEtran}
\IEEEoverridecommandlockouts

\usepackage{cite}
\usepackage{amsmath,amssymb,amsfonts}
\usepackage{algorithmic}

\usepackage{textcomp}
\def\BibTeX{{\rm B\kern-.05em{\sc i\kern-.025em b}\kern-.08em
    T\kern-.1667em\lower.7ex\hbox{E}\kern-.125emX}}

\usepackage{bbm}
\usepackage{comment}
\usepackage{mathtools}
\usepackage{hyperref}
\usepackage{tabularx}
\usepackage{balance}
\usepackage[algoruled, linesnumbered]{algorithm2e}

\usepackage{multirow}
\usepackage{booktabs}
\usepackage{siunitx}

\usepackage{graphicx}
\usepackage{subcaption}
\usepackage{adjustbox}
\usepackage{tikz} 
\usepackage{array}

\usepackage{amstext}   
\usepackage{amsmath}

\usepackage{bbm}
\usepackage{comment}
\usepackage{mathtools}
\usepackage{hyperref}
\usepackage{tabularx}
\usepackage{balance}
\usepackage[algoruled, linesnumbered]{algorithm2e}

\SetCommentSty{mycommfont}

\NewDocumentCommand{\akul}
{ mO{} }{\textcolor{magenta}{\textsuperscript{\textit{Akul}}\textsf{\textbf{\small[#1]}}}}

\usepackage{pgfplots}
\usepackage{subcaption}
\usepackage{xcolor, pifont}

\newcommand{\bates}[1]{{\color{red}#1}}

\newcommand{\Sys}{{\sc Orchid}\xspace}

\newcommand{\Hids}{{Prov-IDS}\xspace}
\newcommand{\hids}{{\Hids}\xspace}

\NewDocumentCommand{\jason}
{ mO{} }{\textcolor{blue}{\textsuperscript{\textit{Jason}}\textsf{\textbf{\small[#1]}}}}
\usepackage[normalem]{ulem} 

\begin{document}\sloppy

\title{ORCHID: Streaming Threat Detection over Versioned Provenance Graphs}

\author{\IEEEauthorblockN{Akul Goyal}
\IEEEauthorblockA{Computer Science\\
University of Illinois at\\
    Urbana-Champaign\\
akulg2@illinois.edu}
\and
\IEEEauthorblockN{Jason Liu}
\IEEEauthorblockA{Computer Science\\
University of Illinois at\\
    Urbana-Champaign\\
jdliu2@illinois.edu}
\and
\IEEEauthorblockN{Gang Wang}
\IEEEauthorblockA{Computer Science\\
University of Illinois at\\
    Urbana-Champaign\\
gangw@illinois.edu}
\and
\IEEEauthorblockN{Adam Bates}
\IEEEauthorblockA{Computer Science\\
University of Illinois at\\
    Urbana-Champaign\\
batesa@illinois.edu}}

\maketitle

\begin{abstract}
\input{abstract}
\end{abstract}

\input{imgs/figure2} 
\input{introduction}

\input{motivation}
\input{methodology}

\input{results}
\input{discussion}

\section{Related Works}
\label{sec:related}
\input{related_works}

\input{conclusion}

\bibliographystyle{IEEEtran}
\bibliography{graph, bates-bib-master}

\input{appendix}

\end{document}

%% file: abstract.tex
While Endpoint Detection and Response (EDR)
  are able to efficiently monitor threats by comparing
  static rules to the event stream,
  their inability to incorporate past system context
  leads to high rates of false alarms.
Recent work has demonstrated Provenance-based Intrusion Detection Systems (Prov-IDS)
  that can examine the causal relationships between abnormal behaviors
  to improve threat classification.
However, employing these \Hids in practical settings remains difficult --
  state-of-the-art neural network based systems are only fast in a fully offline
  deployment model that increases attacker dwell time,
  while simultaneously using simplified and less accurate provenance graphs
  to reduce memory consumption.
Thus, today's \Hids cannot operate effectively in the real-time streaming setting
 required for commercial EDR viability.

This work presents the design and implementation of \Sys,
  a novel \hids that performs fine-grained detection of process-level threats
  over a real time event stream.
\Sys takes advantage of the unique immutable properties
  of a versioned provenance graphs to iteratively embed the entire graph
  in a sequential RNN model while only consuming a fraction of the computation
  and memory costs.
We evaluate \Sys on four public datasets, including DARPA TC,
  to show that \Sys can provide competitive classification performance
  while eliminating detection lag and reducing memory consumption
  by two orders of magnitude.


%% file: imgs/figure2.tex
\begin{figure*}[t!]
  \captionsetup{width=\textwidth}
  \centering 
	\includegraphics[width=1\linewidth]{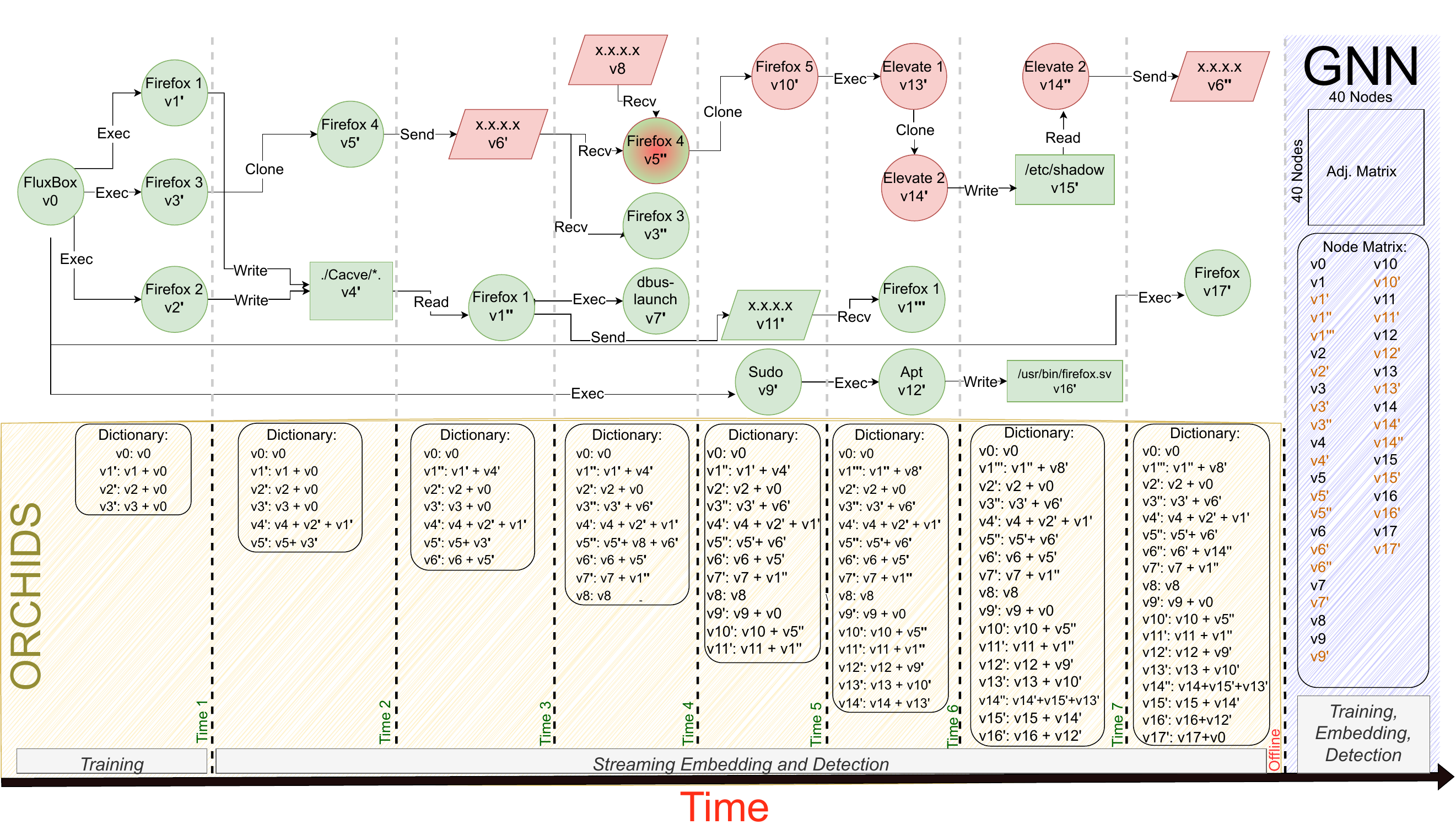}
	\caption{Overview of the \Sys architecture, as demonstrated on an example provenance graph. Green and red nodes represent benign and attack entities, respectively. To create a more precise and acyclic provenance graph, entities are {\it versioned} in the graph as they are updated, e.g., h1' is a new version of h1. In a GNN, the entire graph including testing data must be collected before training begins. In contrast, \Sys continuously embeds and classifies new events as they occur following a preliminary training period. Times 1-7 demonstrate how \Sys's {\it streaming} dictionary has evolved at discrete points in the timeseries.
By only maintaining state on the current version of system entities, \Sys is able to maintain a smaller memory footprint, as can be seen by comparing the dimensions of its dictionary to the GNN's matrices.}

	  \label{img:example:graph}
\vspace{-0.95em}
\end{figure*}

%% file: introduction.tex
\section{Introduction}
Effective logging plays a crucial role in defending against malicious behavior. Most of today's cyber-attacks can forensically be captured through a combination of multiple log streams. However, the sheer volume of logs generated poses a significant challenge for organizations, demanding substantial computational and memory resources for proper analysis and storage. For instance, even in a small data center, daily host logs can quickly accumulate hundreds of gigabytes of data \cite{huber2017syscall}. Moreover, reliant defense against modern cyber-attacks necessitates security tools to offer real-time detection capabilities without excessively burdening the underlying system resources.
Consequently, current commercial security solutions meet these demands by adopting a strategy of lightweight analysis, using pattern matching based on pre-defined rules, implemented on coarse-grained logging systems like Sysmon \cite{Sysmon} or Carbon Black \cite{CarbonBlack}, in place of the heavier analysis of the finer-grained logging systems such as Linux AuditD \cite{auditD}. While this approach seeks to balance performance and system scrutiny, it also introduces a trade-off that exposes organizations to undetected attack steps when certain logs are not collected and analyzed. As such, finding the optimal compromise between tool efficiency and the thoroughness of log data remains a critical challenge in fortifying cyber defenses.


Data provenance has emerged as a potential contender to address some of these challenges, utilizing the low-level system calls to learn malicious behaviors and cutting down on false positives while still identifying malicious behavior. Provenance research timestamps system calls within audit logs to create causal dependency graphs that correctly model system activity. The graphical structure maintains the causal relationship between events regardless of how temporally apart they occur. Much of the current Provenance-based Intrusion Detection Systems (\Hids) research has primarily focused on analyzing these graphical structures to identify the occurrence of anomalous substructures accurately. Initial research \cite{mma2016, hpb+2020, whl+2020, yangprographer, xie2018pagoda} focused on building lightweight systems that provided global classification over the entire graph. However, recent work \cite{ghb+2023, mukherjee2023sec} has shown that these systems were vulnerable to mimicry attacks that allowed the attacker to evade detection. More modern solutions \cite{zengy2022shadewatcher} have adapted to the mimicry problem by focusing on more sophisticated models that provide finer-grained classification (node-level, edge-level). Nonetheless, these solutions have occurred at the expense of creating memory and computationally intensive systems -- analyzing a canonical fully versioned provenance graph calls for \textbf{143.7} GB of memory. These \Hids rely on static audit logs requiring them to be deployed at specific "checkpoints" to transform the input logs into an in-memory graphical format, which the \Hids then uses to train, embed, and detect for abnormal activity.

In this paper, we introduce \Sys (Online Root Cause Host Intrusion Detection System), a stream-based node-classifying Prov-IDS with a low memory footprint and accurate identification of attacker behavior. The foundation for \Sys resides in using a sequential-based machine learning model to model audit logs while adhering to their graphical format. Unlike its predecessors, \Sys does not generate and store the entire provenance graph in memory. Instead, \Sys keeps an $L$-dimensional embedding for each system entity that encodes its full history on the system. This significantly reduces memory usage compared to prior approaches, while the embeddings contain more information than their predecessors. Additionally, to overcome the long-standing challenges with sequential models of capturing long-term dependencies, \Sys utilizes root causes to introduce additional context to the sequential model, allowing for more accurate embeddings. 

In our comprehensive evaluation of \Sys,
  we compare \Sys to GNN-based architectures used in state-of-the-art \Hids,
  including an adaption of GNNs for streaming settings.
We find that \Sys is nearly as effective at threat classification as
  a GNN in an offline model, but significantly more effective
  than GNNs in an online streaming model (e.g., $0.91$ vs. $0.73$ AUC
  on DARPA TC3 Theia \cite{darpa18github}).
We go on to demonstrate that \Sys has a dramatically lower memory footprint
  than GNNs ($2.7$ GB vs. $143.7$ GB on DARPA TC3 Trace \cite{darpa18github}).
Finally, we demonstrate that while \Sys is able to embed and classify
  new events in near real time (0.002 seconds per event),
  {\it in practice offline GNN models may experience detection lag of $18$ to $42$ hours.}


The rest of this paper is organized as follows. In Section 2, we motivate our approach and consider the limitations of prior work. Section 3 presents the threat model that guides our design. We present the design of \Sys in Section 4. Section 5 presents our experiments and performance results. We revisit related work in Section 6, consider limitations of our approach in Section 7, and conclude in Section 8.

%% file: motivation.tex

\section{Motivation}
\label{sec:motiv}

To illustrate the challenges faced Provenance-based Intrusion Detection System (\Hids), we draw upon a realistic example of an Advanced Persistent Threat (APT) attack observed in the DARPA Transparent Computing Attack Engagement 5. The attacker starts off by strategically placing a malicious ad server on a trusted website,
specifically {\tt www.allstate.com}, and waiting for a potential victim with a vulnerable Firefox browser to visit.
Upon connection, the attacker installs a Drakon implant in the memory of the victim's Firefox process resulting in the exploited Firefox establishing a connection back to the attacker's server with the IP address {\tt 189.141.204.211} ready to receive commands. With control of the process, the attacker escalates their privileges on the system by capitalizing on a previously installed driver named ``BinFmt Elevate.''
This driver allows the attacker to execute an Elevate process with elevated root access to the victim system.
The attacker uses the higher privilege process to access sensitive data  {\tt etc/shadow} and subsequently exfiltrate this data back to their server. During the attack, the victim performs benign activities, including updating Firefox.

A causal graph of the attack is depicted in Figure \ref{img:example:graph},
  with red nodes indicating malicious behavior and green nodes representing legitimate
  activity.
The W3C definition of a provenance graph \cite{w3c} states a causal graph must be directed and acyclic.
  We achieve this by employing {\it graph versioning} \cite{mhb+2006} wherein a new node is created for a system entity when it {\it receives} a new information flow and the its ``old'' node becomes {\it locked}.
  
New edges are appended only to unlocked nodes making locked portions of the graph {\it immutable}. Figure \ref{img:example:versioning} (Appendix) visualizes an audit log and its corresponding versioned provenance graph.

\subsection{Key Limitations of Prior Work}
In Figure \ref{img:example:graph}, the attacker's behavior appears noticeably distinct from the normal activities;
  however, in practice provenance-based anomaly detectors face many challenges.
To demonstrate, we consider an exemplar Graph Neural Network (GNN) based \Hids architecture
  used by systems like ShadeWatcher \cite{zengy2022shadewatcher}
  and R-CAID \cite{gwb2024}.
These system represent the state-of-the-art in terms of their ability to perform
  fine-grained entity-level detection in a provenance graph;
  coarser-grained graph-level \Hids are discussed in Section \ref{sec:related}. Note the right hand side of Figure \ref{img:example:graph} visualizes a GNN's internal representation of the Drakon attack.

\paragraph{Memory Overhead} Provenance graphs quickly grow massively in size,
  e.g., $130$GB for a single day on a single machine \cite{mzk+2018}.
This necessitates careful design when choosing how to store and analyze provenance graph.
Ideally, as shown in Figure \ref{img:example:graph}, the GNN would the track state for all $40$ nodes
  within the versioned provenance graph, despite only $17$ system entities existing.  
However, to be more memory efficient, prior work reduces memory and analysis costs by
  using a simplified causal graph that contains cycles.\footnote{The simplified GNN state model used in prior work would contain single node per system entity and the adjacency matrix would allow cycles.}
Unfortunately, this compromise sacrifices temporal ordering and event occurrence information that
  may be crucial to detection.    
For example, benign {\tt Firefox 1} uses an exec call to launch {\tt dbus-launch}
  {\em before} connecting to an IP address.
Conversely, in the attack subgraph, {\tt Firefox 4}
  executes the malicious {\tt Elevate} process
  {\em after} connecting to the malicious IP address.
Loss of temporal information renders both activities almost identical,
  making it harder for \Hids to detect.

\paragraph{Detection Lag} Minimizing an attacker ``dwell time'' \cite{crowdstrike_dwell} by reducing the time to detection is an essential goal for any threat detection system.
However, GNN-based architectures assume access to a {\it static graph} which provides access to
  all training and testing data at once.
This means GNN-based \Hids are required to operate inn
  a {\it fully offline} model where training cannot not begin until
  {\it after} the desired inference period has already occurred.
This is depicted in Figure \ref{img:example:graph}, where the time axis
  shows that the GNN does not begin training until after the attack has concluded.  
This approach is a boon to the attacker, guaranteeing them additional dwell time
  equivalent to the combined lengths of the GNN's costly training and inference periods.

\subsection{Our Approach}


We have shown that effective \Hids design requires operating on {\it lossless} provenance graphs
  while maintaining the memory and computational efficiency for {\it real-time analysis}
  of the event stream.
This requires rethinking the underlying architecture to construct the graph and facilitate
  the embedding and detection process.
\Sys uses an sequential-based neural network (SNN) to learn an embedding function
  on a preliminary portion of the provenance graph (Time 1 in Figure \ref{img:example:graph}).
It then performs real-time analysis of the audit event stream.   
As new events occur, \Sys efficiently embeds and tests the current version for each system entity.
Because \Sys maintains an entry only for the most recent version of each system entity
  and no edge state, the complexity of its state model is dramatically
  streamlined compared to GNN architectures.

To better understand why \Sys can effectively and securely operate under a simpler state model, we will utilize the attack inn Figure \ref{img:example:graph}. Consider at Time $7$ the introduction of malicious {\tt Elevate 2} process (or $v14'$) whose representation is defined as $v14' =v14 + v13'$, where $v14$ is {\tt Elevate 2} feature vector and $v13'$ is an recursive embedding of event history including the vulnerable {\tt Firefox 4} process getting exploited by the attacker. So at Time $8$, when {\tt Elevate 2} reads from {\tt /etc/shadow}, the inherent abnormality of information flow between a highly sensitive data source and a node whose history includes an outside IP address (v6, v8) is represented in the summation of $v15$ and $v14'$. As a result, the $L$-dimensional embedding $v14''$ should appear as an outlier when compared against a model capturing normal behavior. This recurrent structure employed by ORCHID allows for simple computation to create $v14''$ which then fully capture {\tt Elevate 2}'s history without requiring previous versions ($v14', v14$). This reduces the computational and memory costs associated with the detection system while still allowing it to remain accurate. In the following sections we will detail exactly how the architecture of \Sys allows it effectively stream over the incoming event log.

%% file: methodology.tex
\section{Threat Model}
In this work, we consider a threat model that is consistent with past work \cite{mma2016, whl+2020, hpb+2020, xie2018pagoda, zengy2022shadewatcher, yangprographer, hyp+2021}.
We consider an adversary that employs a sophisticated attack strategy,
  drawing from various techniques described in the MITRE ATT\&CK framework \cite{mitre}.
For example, the attacker may engage in mimicry attacks or other forms of  
  masquerading (T1036 \cite{masquerading}), manipulating features of their
  own artifacts to appear legitimate to security tools.
Consistent with prior work, we place the following restrictions
  on the adversary.
We assume the attacker cannot tamper with or turn off mechanisms in the
  trusted computing base of the system, compromised of the operating system,  
  auditing subsystem, and the Prov-IDS.
Beyond these assumptions, we place no further restrictions on the adversary.

\section{\Sys Design}
\label{sec:meth}
In this section, we will detail the design of \Sys. We will start by introducing notation and background knowledge that we will utilize throughout the rest of the paper. Next, we will detail the foundations of \Sys and its use of a sequential-based neural network (SNN) to analyze provenance graphs. We will then describe how \Sys can provide meaningful representations, maintaining a low memory and computational overhead in real-time. Finally, we will explain how \Sys classifies anomalies and how we trained and tested \Sys. 

\input{imgs/figure12}

\subsection{Preliminaries and Background}
\label{sec:meth:prelim}
For a given host system $S$, let the audit log $A$ be a file that continuously records every interaction that utilizes the operating system. Each line $a_i = (v_i, v_j, r_k, t)$ in $A$ represents a directed edge defined as a system call ($r_k$) between the source system entity ($v_s$) and the destination system entity ($v_j$) that occurs on the system at time $t$.
We focus on three different types of system entities (process, file, socket)
  and five different event types.
Table~\ref{fig:table:example} summarizes the system calls
  and provides a graphical representation - importantly, each pair of system entities and system call is unique such that for any two system calls $(v_i, v_j, rel_k)$, $(v_j, v_i, rel_l)$, it is always true that $rel_k \neq rel_l$. Intuitively, the direction of each system call defines the flow of information within the system, and the same system call cannot represent two different flows.

\paragraph{Provenance Graph}
A provenance graph $G = <V,E>$ is a directed acyclic graphical representation of the audit log $A$ adhering to a strict temporal property \cite{w3c}. This property ensures that for every node $v_i \in V$, all incoming edges ${(v_j, v_i, r_k, t_1), \cdots}$ occur chronologically prior to any outgoing edges ${(v_i, v_l, r_m, t_2), \cdots}$.  Previous methods employ a technique known as \emph{versioning} \cite{mhb+2006} to achieve temporal correctness. For a given vertex $v_i$, versioning inserts an identical (same initial feature vector) vertex $v_i'$ whenever an incoming edge occurs after an outgoing edge. For instance, in Figure~\ref{img:example:versioning} of the Appendix, lines $2$ and $6$ in the audit log have ${\tt firefox 1}$ writing and then reading from ${\tt ~/.cache/*}$. In a non-versioned graph, this interaction would be a cycle between ${\tt firefox 1}$ and ${\tt ~/.cache/*}$ violating the temporal property of provenance. The cycle is broken by versioning ${\tt firefox 1'}$, resulting in a provenance graph that is both directed and acyclic \cite{w3c}. Note that once a node ($v_i$) is versioned $v_i'$, its initial state ($v_i$) is immutable and unable to be modified within the graph. Finally, a \emph{causal subgraph} $P(v_d, t)$ is the subgraph resulting from a backward trace from node $v_d$ to all the roots within $G$ at timestep $t$. In a streaming setting, given graph $G$ is constantly updated and $v_d$ has yet to be versioned, $P(v_d, t)$ at timestep $t$ can be different from $P(v_d, p)$ at a later timestep $p$.

\paragraph{Sequence-Based ML and RNN}
Sequential-based machine learning models represent
  the temporal dynamics and inherent dependencies found between data points to generate accurate embeddings for time series. To illustrate, consider a sequential model $f$ operating on a data sequence $s = (s_0, \cdots, s_n)$ and its corresponding sequence of feature vectors $x = (x_0, \cdots, x_n)$ where each token $x_i$ represents a numerical representation of its corresponding element $s_i$. During prediction, the sequential model $f$ takes the input sequence $x$ and produces an output sequence $x' = ({x'}_0, \cdots, {x'}_n)$  where each element ${x'}_i$ captures the relationship between $x_i$ and the other elements in the sequence.
  Past research has introduced sequential-based neural network, including recurrent neural networks (RNNs), transformers, convolutional neural networks (CNNs), hidden Markov models (HMMs), and conditional random fields (CRFs).

  In our work, we leverage RNNs \cite{rumelhart1985learning} due to their use of recurrence to capture temporal dependencies. 
 Consider $f$ to be an RNN; at each step $t$ within $x$, $f$ takes in as inputs both the current feature vector $x_t$ and the hidden state $h_{t-1}$ from the previous timestep.At timestep $t$, $h_{t} = f(x_t, h_{t-1})$ represents the embedding process aggregating the $L$-dimensional feature vector $x_t$ and the previous hidden state $h_{t-1}$ produce a new hidden state $h_{t}$. The updated hidden state $h_{t}$ represents an embedding of the sequence up to timestep $t$.

\subsection{Key Idea}
\label{sec:meth:overview}

Consider a dynamic audit log $A$ where at each timestep $t$, a system call $a_t = (v_i, v_j, r_k, t)$ between the system entities $v_i$ and $v_j$ is added. 

 \Sys \emph{directly} operates on $A$ to construct embeddings more specifically, as each event $a_t = (v_i, v_j, r_k, t)$ occurs, \Sys uses a sequential-based neural network $f$ to generate an embedding using:
\begin{equation}
\label{eq:1}
D[v_j] = f(D[v_j], D[v_i])
\end{equation}
where $D$ is an internal dictionary that maps each vertex to its most up-to-date embedding, and $f$ is the RNN. This procedure is highly {\em localized},
  where each event $a_t$ only updates a single dictionary entry,
  resulting in a constant time operation.

  Equation \ref{eq:1} allows for a system entity $v_j$'s
  entire provenance history $P(v_j, t)$ to be expressed as a single
  feature vector $D[v_j]$.


\subsection{Vectorizing System Entities}
\label{sec:meth:feat}
Generating a numerical feature vector for each system entity within the audit log is necessary for RNNs as inputs.
To accomplish this,
  we first represent each system entity in terms of its full system path,
  which denotes the entity's location on disk.
File paths represent files,    
  processes by the path to their binary executable,
  and sockets by their connecting IP addresses.
We use Doc2Vec \cite{le2014distributed} to generate a $L$-dimensional embeddings for each system entity.

\subsection{Embedding and Detection}
To detect and identify abnormal behavior, \Sys creates a set of benign embeddings representing normal behavior. \Sys achieves by processing the entire training dataset and generating an $L$-dimensional embedding for each system entity.

With this set of training embeddings, \Sys can detect anomalies in unseen data. For each new edge $a_t = (v_i, r_k, v_j)$ streamed in, \Sys generates an $L$-dimensional embedding for $v_j$ using Eqn.~\ref{eq:1} that is then compared against the set of training embeddings. The distance between $v_j$'s embedding and its closest neighbor determines its anomaly score. If this distance exceeds a pre-defined threshold $\alpha$, the \Sys flags $v_j$ as anomalous.


\subsection{Accounting for Long Term Dependencies}
\label{sec:meth:gradients}


Sequential based neural networks have long been recorded \cite{hochreiter1997long} to have issues modeling long sequence lengths. 
To harden \Sys and allow it to capture long-term dependencies,
  we propose the introduction of ``root-node'' embeddings. 

Root causes are widely acknowledged as a vital source
of information indicating the attacker's origin. Moreover, unlike later attack steps, the attacker cannot manipulate root causes
without directly tampering with the log.
 We incorporate root causes  by introducing an additional component called "root embedding." We reflect this component in a modification to the RNN's update function such that:  
\begin{equation}
\label{root}
h_i = w*(h_{i-1}) + b*(x_i) + c*[\frac{1}{n}\sum_{i=0}^{n}\{r_i\}]
\end{equation}
where $\{r_i\}$ is the set of root nodes associated with element $i$ in the sequence, and $c$ is a learnable model weight-optimized to balance the information introduced by root embedding. Root embedding maintains significant relationships over extended distances, invisible to the RNN, thereby enhancing \Sys' resilience against an active attacker.

%% file: imgs/figure12.tex
\begin{table}[t]
  \captionsetup{width=\columnwidth}
  \centering 
	\includegraphics[height=3cm]{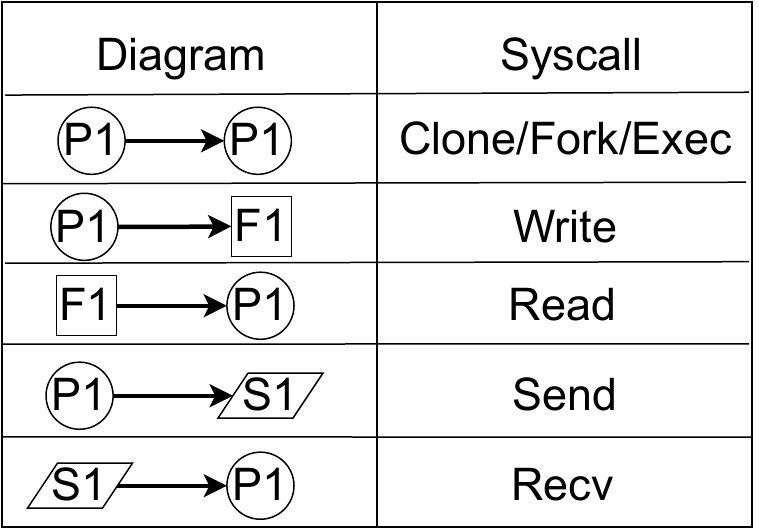}
	\caption{Table visualizing the graphical representation for the $5$ different event types \Sys focuses on.}
	  \label{fig:table:example}
\end{table}

%% file: results.tex
\section{Evaluation}
In this section we provide results evaluating \Sys. For each experiment we wanted to answer the following research questions:

\begin{itemize}
    \item \emph{RQ1} What are the relative contributions of \Sys' different components and parameters to classification performance?
    \item \emph{RQ2} How accurate is \Sys in identifying anomalous behavior as compared to past approaches?
    \item \emph{RQ3} What are the performance gains and benefits that are accrued by \Sys over previous work?
 \end{itemize} 


\subsection{Datasets}
To evaluate the performance of \Sys we use $4$ different datasets: StreamSpot, DARPA Transparent Computing Engagement 3 Theia and Trace, and ATLASv2.

\paragraph{StreamSpot} is a multi-graph dataset describing five benign web browsing behaviors (CNN, YouTube, video game, GMail, file download)
  and one attack behavior (Firefox drive-by-download exploit).
  Each behavior is automated and repeated $100$ times, resulting in $600$ total graphs,
  where each graph is relatively small ($50K$ edges to $700K$ edges).
As stated in Section~\ref{sec:meth}, \Sys's design intrinsically supports multiple disconnected graphs without issue,
  So, we combined all the graphs into a single audit log with unique entity identifiers for each graph.
We interleave the five benign behaviors and then append all the attack behaviors at the end.
We permit \Sys to train on a single instance of each benign behavior for the preliminary training period.
As a result, \Sys training, validation, and testing split is $1\%$, $1\%$, $98\%$ respectively.
 \paragraph{DARPA Transparent Computing  E3} These datasets consist of system graphs collected from 
 various hosts over a $10$-day period, with both benign activity on each host
 and red team attacks against each host \cite{darpa18github}.
 We make use of the Theia (Ubuntu Linux 12.04), comprised of $21M$ edges,
 and Trace (Ubuntu Linux 14.04), comprised of $260M$ edges.
 The majority of attacks performed involved exploiting Firefox to run drakon on the victim host.
 Most attacks were performed towards the end of the $10$-day period, but some attacks happened earlier.
 We restrict the training data to the amount of benign data before the first attack,
   or about $1$ day of data.

\paragraph{ATLASv2}
  replicates the attack engagements conducted by Alsaheel et al. \cite{anm+2021}
  in their evaluation of the ATLAS threat investigation system.
It comprises $10$ attack chains, including both single-host and two-host attacks,
although many of the steps in the chains partially overlap between attacks.
A notable limitation of the original ATLAS dataset was the lack of meaningful
  benign activity. While the authors did include limited benign activity, 
  it does not follow a realistic usage profile and is
  insufficient to train one class anomaly detection models.
ATLASv2 \cite{riddleatlas} addresses these limitations through the introduction of a
  four-day period of benign naturalistic activity generated by human users,
  followed by a final day in which background activity continues as the
  attacks occur.
We make use of the Microsoft Windows Security Auditing logs
  from this dataset, with the first day used for preliminary training.

\paragraph{Ground Truth Labeling}
The absence of a standardized ground truth labeling methodology
  for intrusion detection, datasets present a notable challenge to the community.
Existing datasets employ disparate labeling approaches.
Streamspot labels entire logs/graphs as either malicious or benign,
  the original ATLAS dataset labels only a minority ($2$-$3$) of system entities
  as malicious in each attack chain,
  while DARPA TC datasets provide qualitative ground truth descriptions
  but no explicit labeling at all.
This lack of uniformity in ground truth labeling raises concerns
  of both reproducibility and bias in IDS evaluations.
To avoid these issues,
  we make use of the recently released {\it Recover Every Attack Process} (REAPr)
  label set \cite{reapr-ground-truth}.
REAPr contains process-level ground truth labels for all of the datasets above
  using a standardized provenance-based labeling methodology.
Marking every process on the path between an attack's root causes
  and terminal impacts as malicious, REAPr provides a more comprehensive
  set of attacker-influenced processes while avoiding the potential
  for experimenter bias.
Details of the REAPr labeling methodology can be found online.\footnote{\url{https://bitbucket.org/sts-lab/reapr-ground-truth}}
  
\input{imgs/figure10}

\subsection{Hyperparameter Tuning (RQ1)}
We now explore the relative contributions of
  different components of the \Sys framework.
We consider six facets of \Sys:
  selection of the RNN model,
  number of model layers,
  amount of training data,
  inclusion of root node adaptation,
  size of the feature vector,
  and the classification task.
To facilitate faster testing of various model parameters,
  we use the StreamSpot dataset (except where otherwise noted) due to its smaller size.
The outcomes of the hyper-parameter search are illustrated in Figure \ref{fig:hyper},
  wherein each plot  presents the Receiver Operating Characteristic (ROC) curve
  that compares \Sys' true and false positive rates under variable classification thresholds.

\paragraph{RNN Model Comparison}
As we previously demonstrated in Section~\ref{sec:meth:overview}, a sequential neural network is best adapted to  \Sys because it can effectively model causal paths given limited information about the underlying sequence.
However, as noted in Section~\ref{sec:meth:gradients} the SNN suffers from a memory loss issue,
  which may impede its ability to model longer sequences of attack behaviors effectively.
Cho et al.'s Gated Recurrent Units (GRUs) mechanism \cite{cho2014learning} addresses this limitation of RNNs through a gating mechanism that selectively updates and resets information.
Comparing the RNN and GRU architectures with all other model parameters constant (Fig. \ref{fig:eval:hyper6}),
  the GRU outperforms the RNN even on the relatively short-lived Streamspot dataset ($0.59$ vs. $0.42$ AUC).
We thus employ GRU as the model backbone for \Sys.
  It should be noted that this results in a higher computational demand, which extends training and testing times.

\paragraph{Number of Layers}
Previous research \cite{carlini2022quantifying} has highlighted the impact of model size on performance.
Continuing with the GRU, we vary the number of model layers ($1, 5, 10, 20$) in Figure \ref{fig:eval:hyper1}.
As anticipated, increasing the number of layers improved classification performance,
  but concomitantly increases the computational resources required.
While the $1$-layer GRU performs markedly worse ($0.37$ AUC),
  we observe diminishing returns on increased model size between the $5$-, $10$-, and $20$-layer models.
We chose a $10$-layer GRU for the subsequent experiments.

\paragraph{Root Node Optimization}
To mitigate RNNs' difficulty in modeling longer-term dependencies,
  \Sys includes an optimization that directly embeds every entity's root causes alongside its local context (Section~\ref{sec:meth:gradients}).
We compare the performance of \Sys with and without this optimization in Figure \ref{fig:eval:hyper5}.
Including this optimization improves \Sys' classification overall ($0.59$ vs. $0.54$ AUC),
  pronounced at low FPR values.
These findings demonstrate that the additional context provided by root causes
  can help detect additional attack entities that \Sys might have overlooked.

\paragraph{Feature Vector Size}
The feature vector that serves as a numerical representation of each system entity is a crucial parameter in \Sys.
While richer feature vectors may better capture underlying data distributions,
  the inclusion of additional information could also introduce noise.   
Comparing different vector sizes in Figure \ref{fig:eval:hyper3},
  a vector of size $64$ achieves the most favorable ROC curve ($0.59$ AUC).
We use this size for the remainder of our experiments.

\paragraph{Amount of Training Data}
While the size of \Sys' preliminary training dataset could play a crucial role in its performance,
  moving into the streaming phase as quickly as possible is desirable. To determine an appropriate training dataset size, we make use of the larger Theia dataset.
Starting with a training dataset of $10K$ edges, we gradually increased the size to $1M$ edges, where $1M$ edges correspond to a single day of logging on the host system.
As illustrated in Figure \ref{fig:eval:hyper2},
  there is a general trend of improved classification performance as training data increases.
We choose to train \Sys on $1$M events ($0.91$ AUC) in subsequent experiments;
  this amount is roughly proportional to a full day of activity and is likely to generalize better
  to other datasets.

\paragraph{Classification Task}
While \Sys can classify any system entity in the provenance graph, performance could vary between process entities and data entities like files or sockets.
Continuing with the Theia dataset, we compare process-only to all-entity embedding performance in Figure \ref{fig:eval:hyper4}.
Surprisingly, we do not observe a meaningful difference between the two tasks,
  with ``All Nodes'' classification negligibly outperforming ``Process Only'' ($0.91$ vs $0.90$ AUC).

\input{imgs/figure9}

\subsection{Classification Results (RQ2)}
\label{subsec:classifcation}

We now evaluate the classification performance of \Sys.
Recall every process on the attack path is marked as malicious,
  meaning \Sys is being evaluated on its ability to identify
  {\it every attacker-influenced process} in each dataset.

\paragraph{Comparison Baseline}
Benchmarking \Sys's performance on entity-level classification excludes most prior work, the majority of which does whole-graph classification \cite{mma2016, whl+2020, hpb+2020, xie2018pagoda} and is particularly vulnerable to mimicry attacks \cite{ghb+2023, mukherjee2023sec}.
Even sub-graph systems like SIGL \cite{hyp+2021} and Prographer \cite{yangprographer}
  still, fundamentally perform graph-level classification, but on smaller time slices.
Shadewatcher \cite{zengy2022shadewatcher} is the only GNN-based IDS that performs entity-level classification. Unfortunately, when we contacted the authors of Shadewatcher,
  they informed us they built Shadewatcher on proprietary code that they could not release.

Instead, we evaluate \Sys against two Graph Neural Network architecture variants.
The first model, {\bf \it Full-GNN}, represents a traditional offline deployment model where the model constructs the full graph (training and test data) before training begins.
This model performs nearly identically to ShadeWatcher on the same dataset (Trace) 
  that appeared in their evaluation \cite{zengy2022shadewatcher}.
The second model, {\bf \it Stream-GNN},
  attempts to adapt the Full-GNN to a streaming setting.
The Stream-GNN model uses the same training data
  given to \Sys in each experiment, and is also unable to
  pre-populate its adjacency matrix and node matrix
  to reflect events and entities that only appear during the test period.

Stream-GNN is thus an approximation to the discriminatory power
  of a GNN under \Sys' deployment model.

Another limitation of GNN architectures is their inability to efficiently represent lossless provenance graphs (e.g., versioned).
Unsurprisingly, {\it attempting to train a PyTorch \textbf{GNN}
  on a versioned provenance graph quickly results in an out-of-memory error}
  even on a well-provisioned GPU server.
Provenance graphs, even those that describe a single system,
  are simply much larger than the graphs used to evaluate other graph learning systems,
  and current architectures cannot scale.
As a workaround,
  our baseline models use a simplified provenance representation
  that fails to capture the temporal ordering between events
  and results in {\it false provenance}.
For example, if a process reads from file A, writes to file B, and then reads from file C, a GNN will incorrectly include file C in file B's embedding.
This is the same approach used by previous GNN-based approaches \cite{zengy2022shadewatcher, hyp+2021}.
A hyperparameter search revealed that $2$ layers and $8$ heads delivered the optimal performance for both GNNs.
We trained each GNN for a maximum of $1000$ epochs with early stopping, using Cross-Entropy loss and the Adam optimizer to tune the model weights.

\paragraph{Streamspot Dataset}
Results for Streamspot can be found in Figure \ref{fig:eval:roc1}.
Across the entire continuum of detection thresholds,
  Full-GNN significantly outperforms \Sys ($0.78$ vs. $0.59$ AUC).
Unsurprisingly, Full-GNN creates a more effective embedding model, given its ability to learn on edges after the training period. We expected Full-GNN to represent a performance ceiling for \Sys.

However, in the critical range of detection thresholds with low FPR values,
  \Sys offers several effective performance rates.  
\Sys can detect $23\%$ of all malicious entities, representing at least one detection for each attack behavior,
with $0\%$ FPR, and nearly reaches the Full-GNN's $0\%$ FPR detection rate while admitting just $9\%$ FPR.
On the other hand, given comparable training data (Stream-GNN),
  the GNN's performance hovers around chance ($0.51$ AUC).

\paragraph{Theia Dataset}
In Figure \ref{fig:eval:roc3}, we see a similar trend to StreamSpot where Full-GNN outperforms \Sys ($1.00$ vs. $0.91$ AUC) while \Sys outperforms Stream-GNN ($0.91$ AUC vs. $0.73$ AUC). 

Furthermore, \Sys achieves $97$\% detection with $15\%$ FPR, indicating that \Sys can separate most of the attack within Theia from benign behavior. However, certain attack entities are more challenging to separate as they may only be tangentially related to the attack.

\paragraph{Trace Dataset}
Against the much larger Trace dataset (Figure \ref{fig:eval:roc4}),
  we are surprised to see that \Sys outperforms the Full-GNN (0.88 vs 0.63 AUC).
In the critical region of low FPR detection thresholds, 
  Full-GNN offers a higher detection rate of $35\%$ at $0\%$ FPR,
  while \Sys can offer a $60\%$ TPR with a manageable $12\%$ FPR.
It's difficult to say which is preferable,
  and both systems can detect at least one entity in each attack.
Interestingly, Stream-GNN outperforms Full-GNN across the entire continuum ($0.8$ vs. $0.63$ AUC),
  although its gains come at impractically high FPR values.  

\paragraph{ATLASv2 Dataset}
Results in Figure \ref{fig:eval:roc2} deviate from the previous pattern.
Here,
  \Sys unambiguously outperforms the Full-GNN,
  offering superior TPR rate at $0\%$ FPR and approaching $89\%$ TPR at near-zero FPR.
Stream-GNN's performance again hovers at $0.51$ AUC.
  
\paragraph{Remarks}
For three out of the four datasets,
  \Sys was surpassed in classification performance by Full-GNN.
We expected this, given that the GNN 
  had the advantage of learning its embedding function in a static graph that represented the entire dataset, providing more information to the model during training time.
 Full-GNN could overcome the distribution shifts present between the training and testing datasets,
enabling it to model benign activity more accurately.
Interestingly, \Sys performed better on the ATLASv2 dataset than the Full-GNN.
One potential reason for this outcome is that certain attack nodes in ATLAS were spaced out from each other,
  making them unable to be directly correlated within the limited $K$-hop neighborhood view of the Full-GNN.
Conversely, \Sys,
  could capture longer-term dependencies, allowing it to discern patterns that the full GNN's local viewpoint may miss.
Regardless of performance variance by dataset, both systems
  could detect one or more malicious entities in each attack at a low FPR threshold.  

\input{imgs/figure11}

\subsection{Performance Evaluation (RQ3)}

So far, we have demonstrated that \Sys can accurately classify malicious activity, performing similarly to an offline GNN trained on larger amounts of data.
We now evaluate \Sys runtime performance in a streaming deployment model,
  considering issues of resource consumption and detection latency,
  as compared to prior work.

\paragraph{Memory Consumption}
\Sys operates with significantly reduced memory consumption compared to previous approaches.
By processing events in the graph in a streaming fashion and tracking embeddings for each system entity,
  \Sys reduces the terms in its memory consumption from $|E| + |V|$ to $|V|$.\footnote{Technically, the GNN doesn't have an $|E|$ term 
    because it reserves a space for every possible edge. It is actually $|V|*|V| + |V|$.}
The memory reduction acquired by removing the edge term $|E|$ compounds over time
  as the number of system entities grows slower than the number of system events. Intuitively, activity predominately consists of existing system entities interacting with other system entities rather than generating new system entities. 

\input{tables/table1}

This assertion is validated within Figure \ref{img:memory},
  where we plot \Sys's memory consumption on increasingly larger portions
  of the Trace dataset.
To provide context, we also plot the memory consumption of Stream- and Versioned-GNN. Stream-GNN represents the memory consumption of the GNN on an unversioned graph, ignoring the provenance graph's temporal attributes. The memory consumption of Stream-GNN at the end of the dataset represents the memory consumption for Full-GNN. Versioned-GNN represents the memory consumption of GNN on the more precise {\it versioned} graph that \Sys utilizes, requiring $143.7$GB to store in memory. 
The difference in memory consumption
  between {\it Full-GNN} and {\it Versioned-GNN} underscores the reality
  that GNNs utilize less precise graph representations. 
Even with a less precise graph representation, Stream and Full-GNN maintain a higher memory footprint ($10.4$ GB) than \Sys whose footprint is at most $2.7$ GB while modeling the entirety of the $270$ GB dataset

\input{imgs/figure13}
\paragraph{Computation Cost}
We now report on the raw processing speed of each model,
  inspecting the runtime of three model tasks: preprocessing, training, and embedding and detection.
Preprocessing indicates the time required to map the log into memory and process it into a provenance graph.  
Training is an upfront cost for \Sys and Stream-GNN, but a backend cost for the Full-GNN that immediately precedes embedding and detection; therefore, it is important to note that these {\it runtime costs} do not yet reflect {\it detection latency} for \Sys and Stream-GNN.
An important performance distinction between our RNN architecture and the GNNs is the opportunity for parallelization; while \Sys operates on sequential edges at a constant cost, the GNN can parallelize
  across portions of the static graph.
To reflect this advantage for the Stream-GNN,
  we report embedding and detection for different {\it Checkpoint Intervals} of data.
We report the total time for the Stream-GNN to embed and classify
  1 edge at a time, 1K edges, etc., with EOF indicating the full test batch.  

Results are summarized in Table \ref{tbl:computational}.  
\Sys has an advantage in the pre-processing stage because it does not need to generate
  a provenance graph, and map the edges into memory.
When comparing the offline training time between \Sys and Stream-GNN, Stream-GNN provides a better runtime as it can batch over the static provenance graph, allowing for greater parallelization. While \Sys does not utilize batching, it is possible to batch non-dependent events together, allowing for faster training time.
In comparing the embedding and detection (ED) time for \Sys and Stream-GNN, Stream-GNN outperforms \Sys when operating on batches of $1M$ edges or greater due to the GNN's ability to
  parallelize computation over a known static graph.
In contrast, \Sys must operate on one large stream of sequential data.

{\it Without parallization, the GNN requires  $160$ billion seconds to complete, a $360$K times increase over \Sys.} For the same reasons, \Sys test time costs are independent of checkpoint intervals, processing edges at a constant rate of $300$ edges per second.

\paragraph{Detection Latency}
The true cost of the GNN deployment model is exposed 
  when considering the lag between an event's occurrence and its classification.
Based on the time required to train and test the Full-GNN model on the Trace dataset
  (67.4K seconds, or 18 hours),
  consider a practical deployment model where a GNN classifier trains on
  the previous 14 days' worth of data at midnight each night, then immediately
  attempts to classify the observed events from the previous day.
Instead of re-training and detecting Full-GNN on each of the $14$ days of Trace,
  we averaged over the time Full-GNN took to run over the different datasets
  ($\S$\ref{subsec:classifcation}) to estimate the average time
  it would take Full-GNN to train and test every night. 
We compare the resulting detection lag to \Sys'
  latency in its streaming embedding and detection mode.  
  
The results are pictures in Figure~\ref{fig:lag} for the Trace dataset.
The x-axis depicts the time series of Trace,
  while the y-axis depicts the expected detection lag for an event that
  occurs at that time.
Depending on the time of day, \Sys reduces detection lag by a minimum
  of {\it 10 million times} up to a maximum of {\it 43 million times} by
  eschewing the GNN's offline deployment model.
The minimum time detection lag for Full-GNN
  for an event that occurred directly right before analysis
  would still be $18$ hours, while the maximum time for an event
  that occurred right after analysis occurred would be $42$ hours.   
\Sys, on the other hand,
  could process each event after it occurred, providing a detection lag time of $0.002$ seconds. By streaming, \Sys is able to considerably reduce the attacker's ``dwell'' time on the system, becoming more in line with the runtime associated with commercial EDR tools \cite{gartneredr}.

%% file: imgs/figure10.tex

\begin{figure}[t!]
  \begin{subfigure}[]{0.49\linewidth}
    \centering
    \includegraphics[height=1\linewidth]{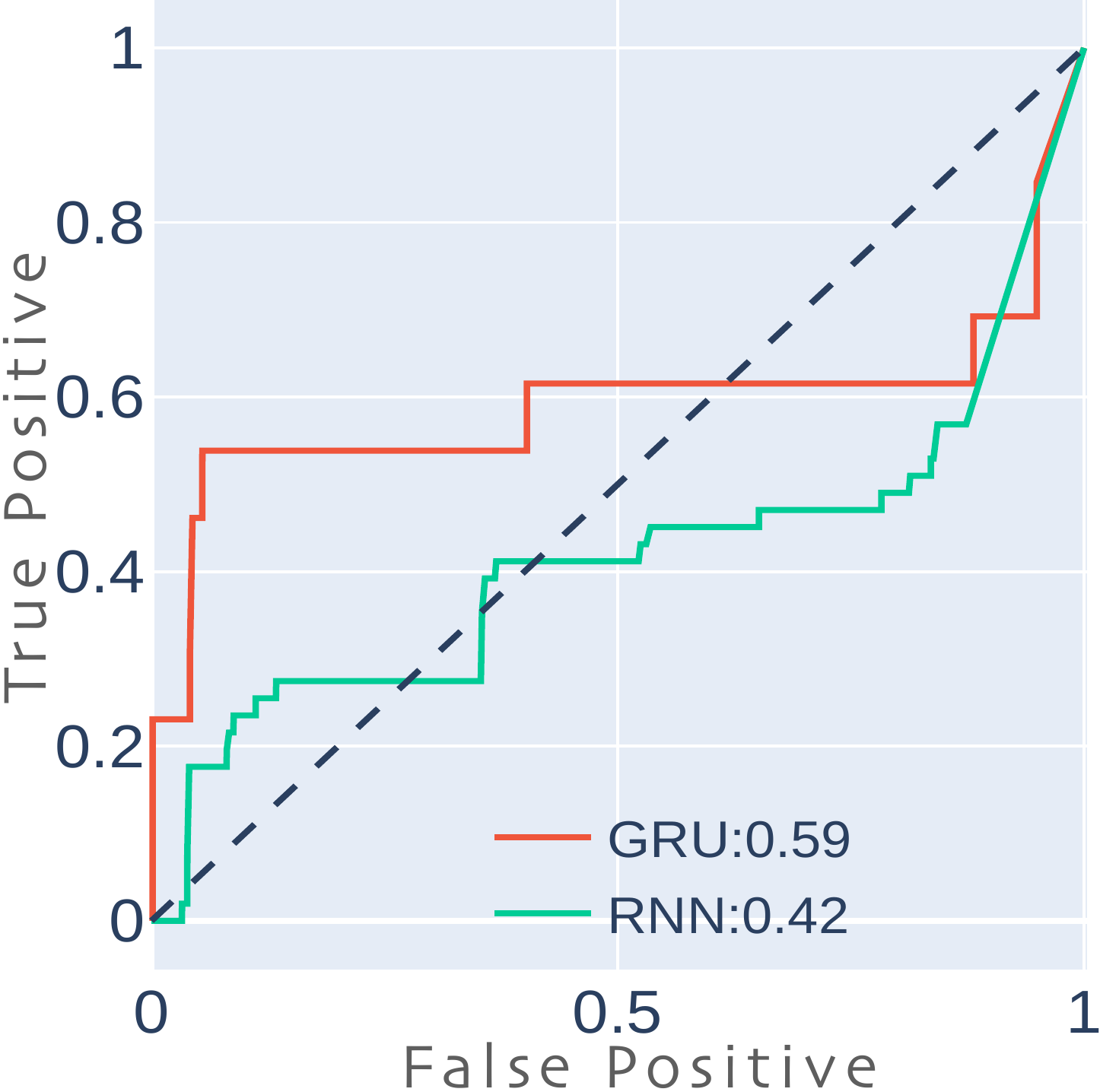}
    \caption{RNN Model Comparison}
    \label{fig:eval:hyper6}
  \end{subfigure}
  \begin{subfigure}[]{0.49\linewidth}
    \centering
    \includegraphics[height=1\linewidth]{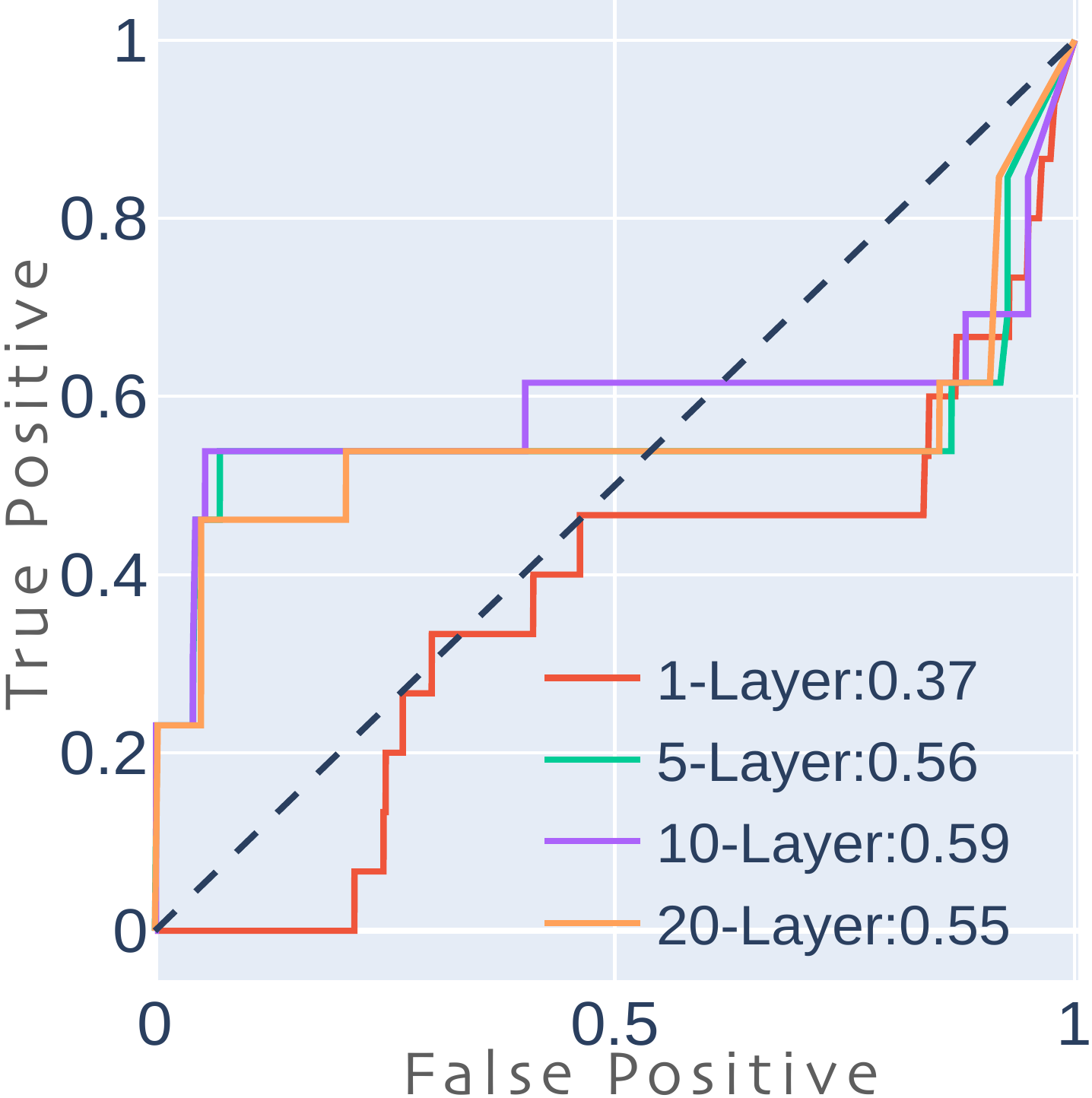}
    \caption{Number of Layers}
    \label{fig:eval:hyper1}
  \end{subfigure}
  \begin{subfigure}[]{0.49\linewidth}
    \centering
    \includegraphics[height=1\linewidth]{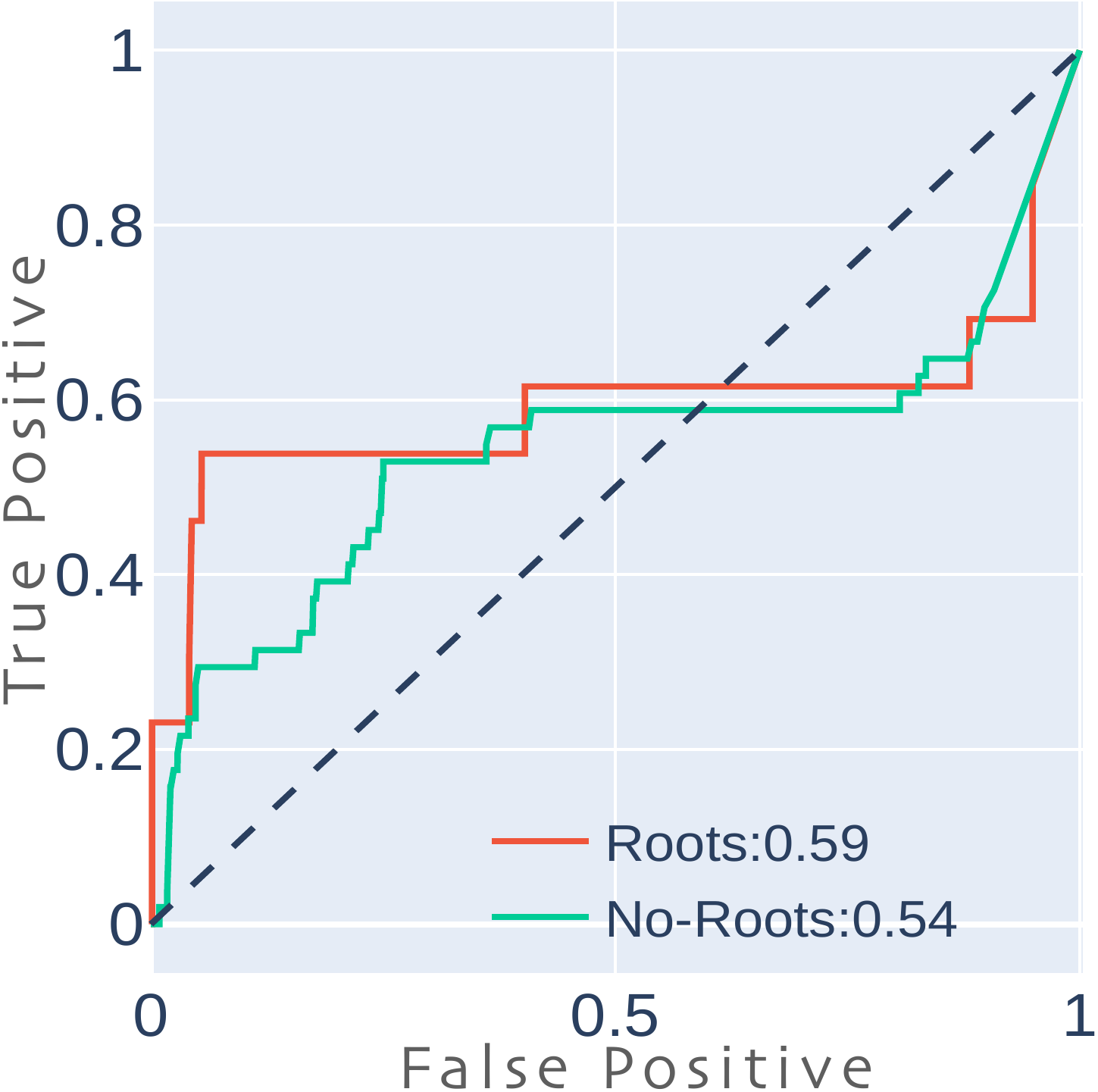}
    \caption{Root Node Optimization}
    \label{fig:eval:hyper5}
  \end{subfigure}
  \begin{subfigure}[]{0.49\linewidth}
    \centering
    \includegraphics[height=1\linewidth]{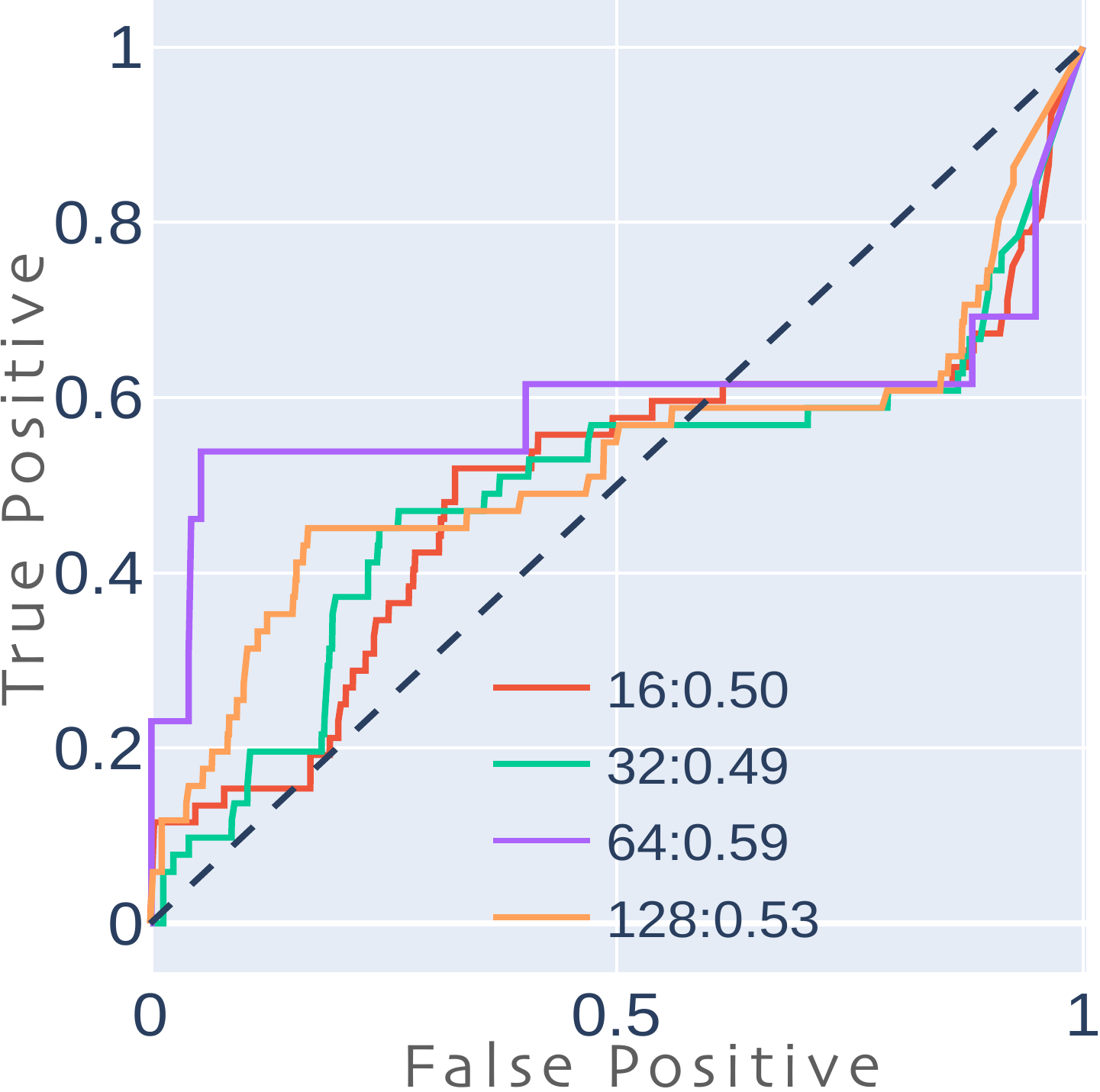}
    \caption{Feature Vector Size}
    \label{fig:eval:hyper3}
  \end{subfigure}
  \begin{subfigure}[]{0.49\linewidth}
    \centering
    \includegraphics[height=1\linewidth]{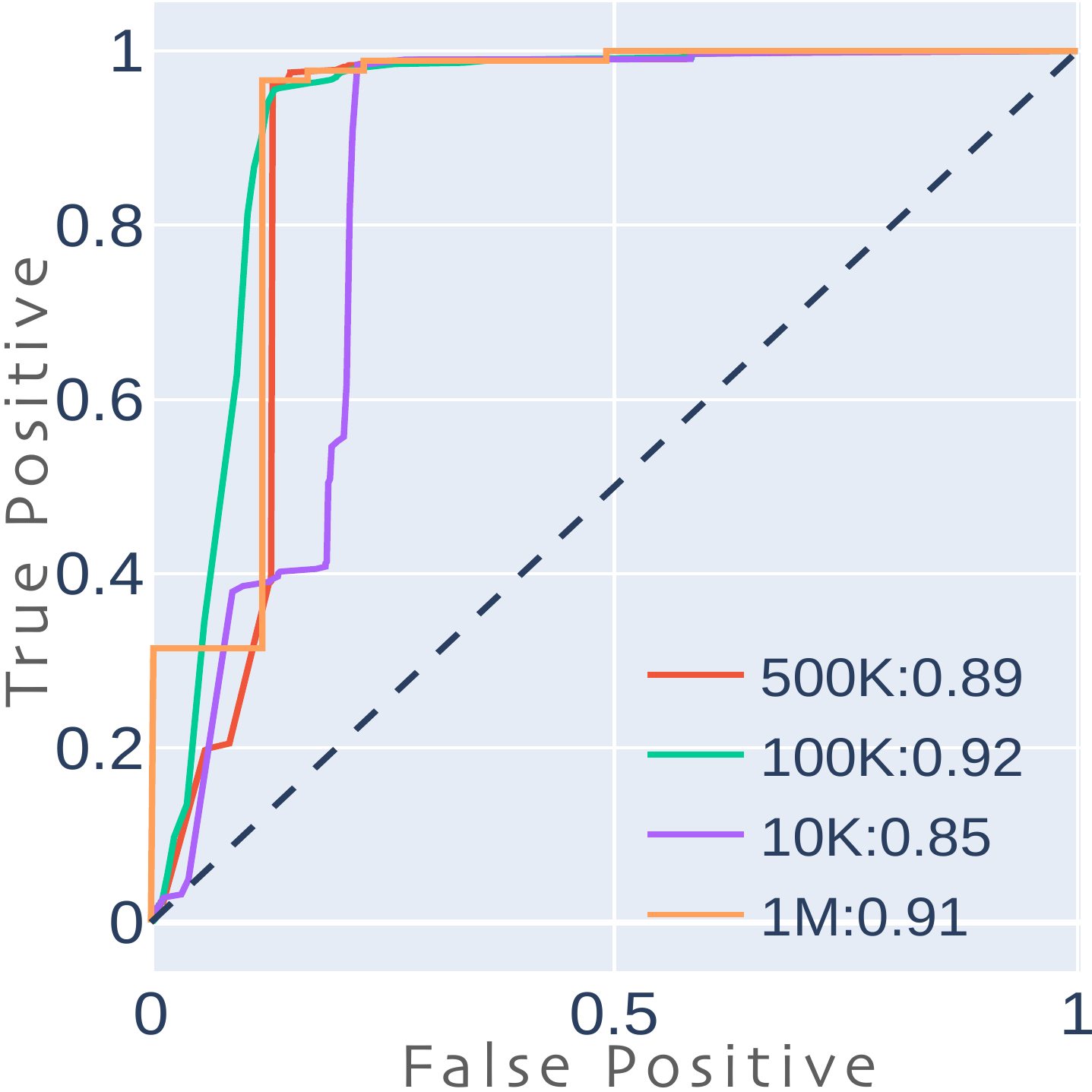}
    \caption{\hspace{-0.5mm}Amt. of Training Data (Theia)}
    \label{fig:eval:hyper2}
  \end{subfigure}
  \begin{subfigure}[]{0.49\linewidth}
    \centering
    \includegraphics[height=1\textwidth]{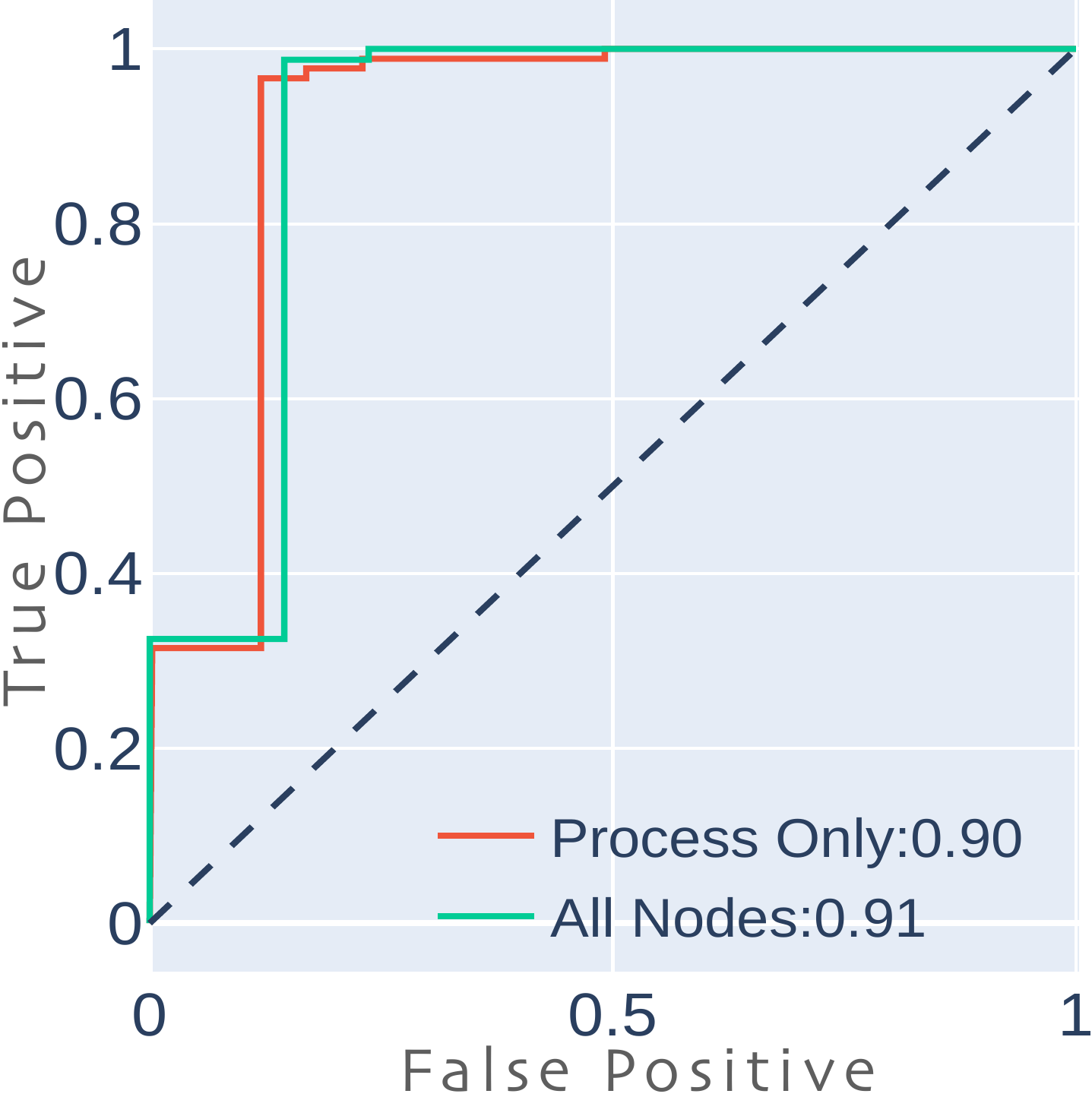}
    \caption{Classification Task (Theia)}
    \label{fig:eval:hyper4}
  \end{subfigure}
  \caption{Hyperparameter tuning of ORCHID. Area Under Curve (AUC) for each line is reported in the legend.}
    \label{fig:hyper}
\end{figure}

%% file: imgs/figure9.tex
\begin{figure*}[t]
  \centering
  \begin{subfigure}[b]{0.24\linewidth}
    \includegraphics[height=\linewidth]{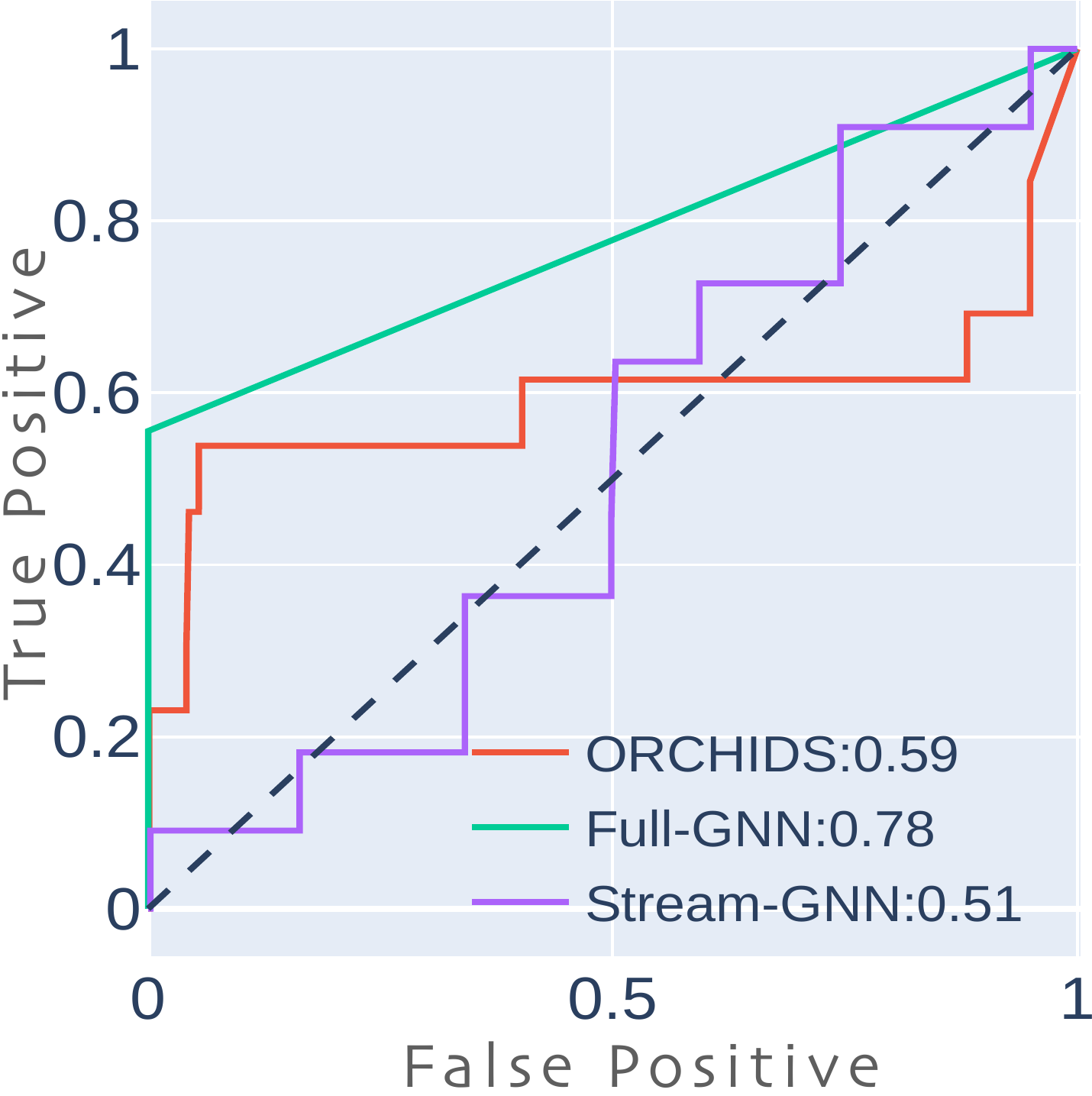}
    \caption{StreamSpot}
    \label{fig:eval:roc1}
  \end{subfigure}\hfill
  \begin{subfigure}[b]{0.24\linewidth}
    \includegraphics[height=\linewidth]{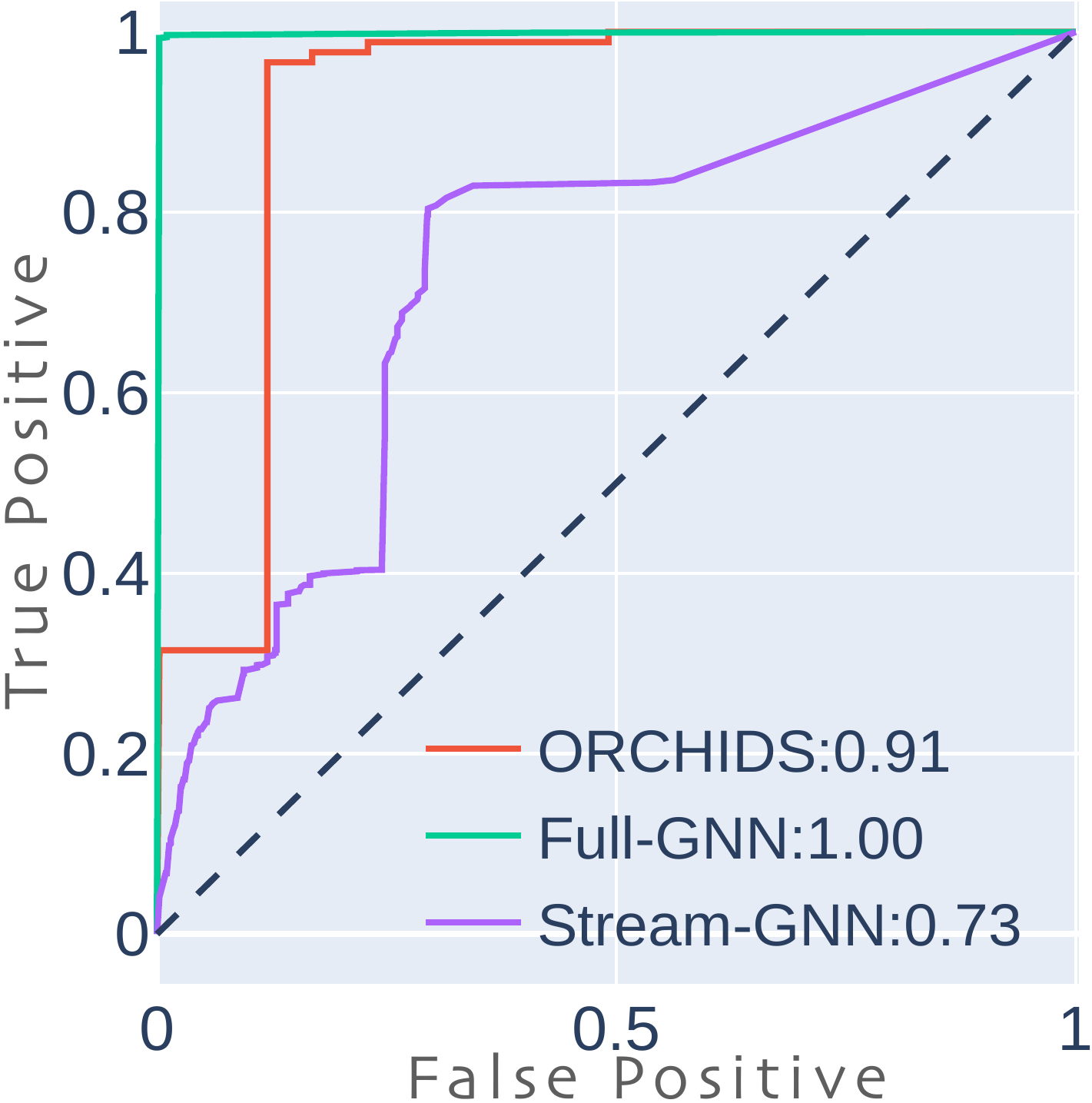}
    \caption{Theia}
    \label{fig:eval:roc3}
  \end{subfigure}\hfill
  \begin{subfigure}[b]{0.24\linewidth}
    \includegraphics[height=\linewidth]{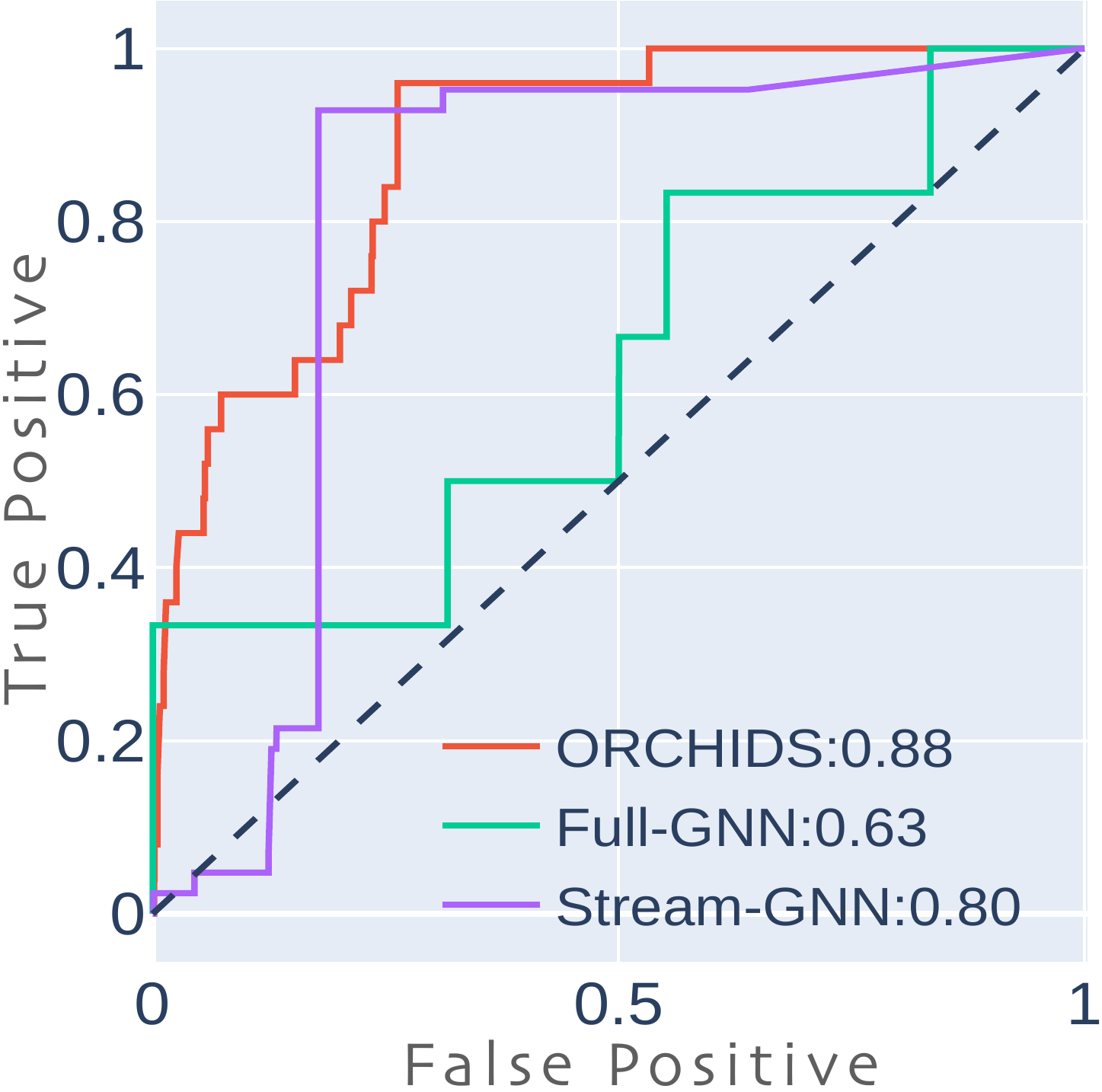}
    \caption{Trace}
    \label{fig:eval:roc4}
  \end{subfigure}\hfill
  \begin{subfigure}[b]{0.24\linewidth}
    \includegraphics[height=\linewidth]{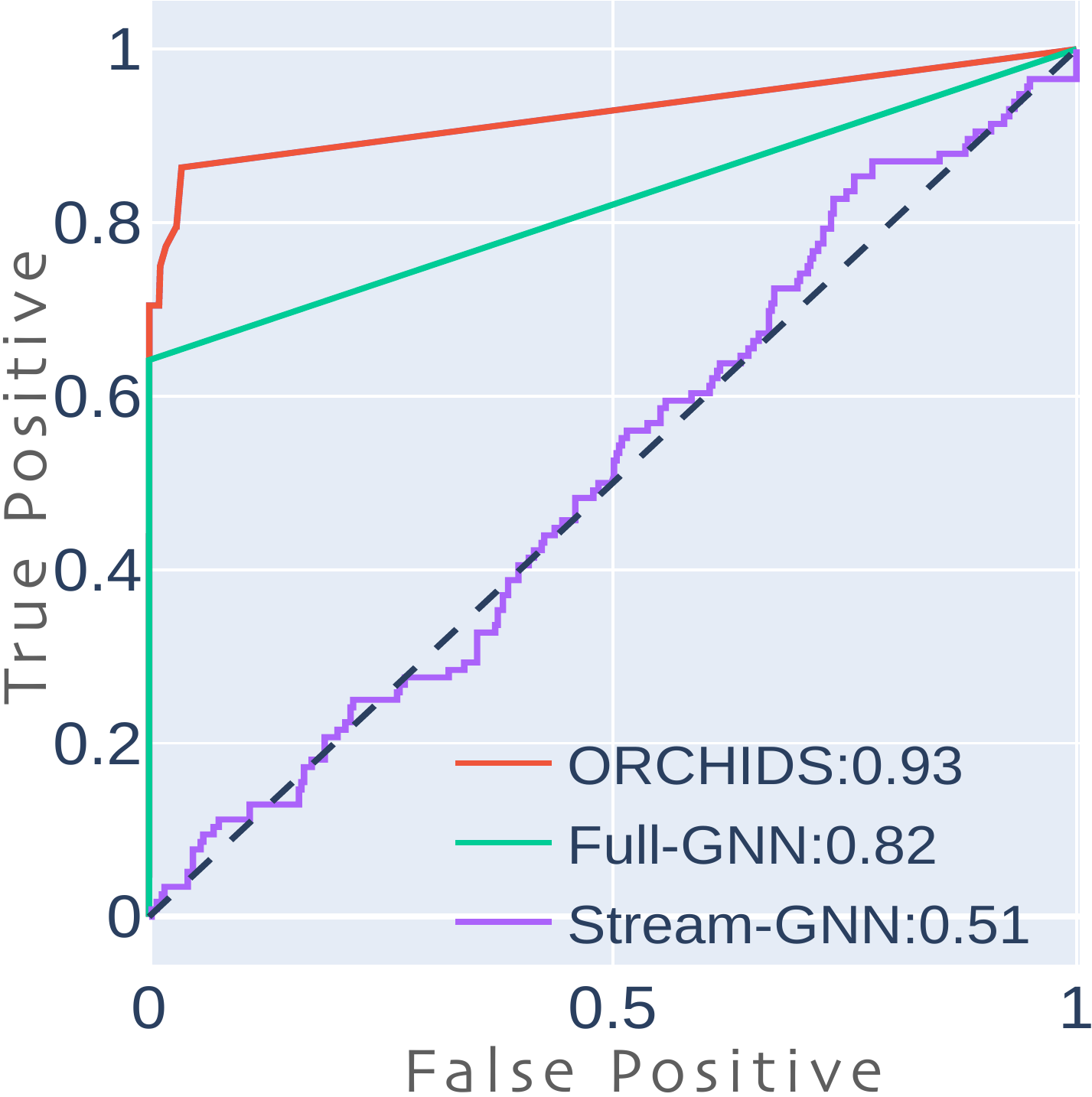}
    \caption{ATLAS V2}
    \label{fig:eval:roc2}
  \end{subfigure}
  \caption{Performance of ORCHID, reported as ROC curves, as compared to GNN-based approaches. In a streaming setting, \Sys' performance is  generally comparable to an offline GNN deployment (Full-GNN), particularly in the low-FPR regions of the plot that denote plausible detection thresholds. Adapting the GNN to an online setting with equivalent training data to \Sys (Stream-GNN), the GNN model is thoroughly outperformed. 
  }
  \label{fig:roc1}
\end{figure*}

%% file: imgs/figure11.tex
\begin{figure}[t]
  \captionsetup{width=\columnwidth}
  \centering 
	\includegraphics[height=5cm]{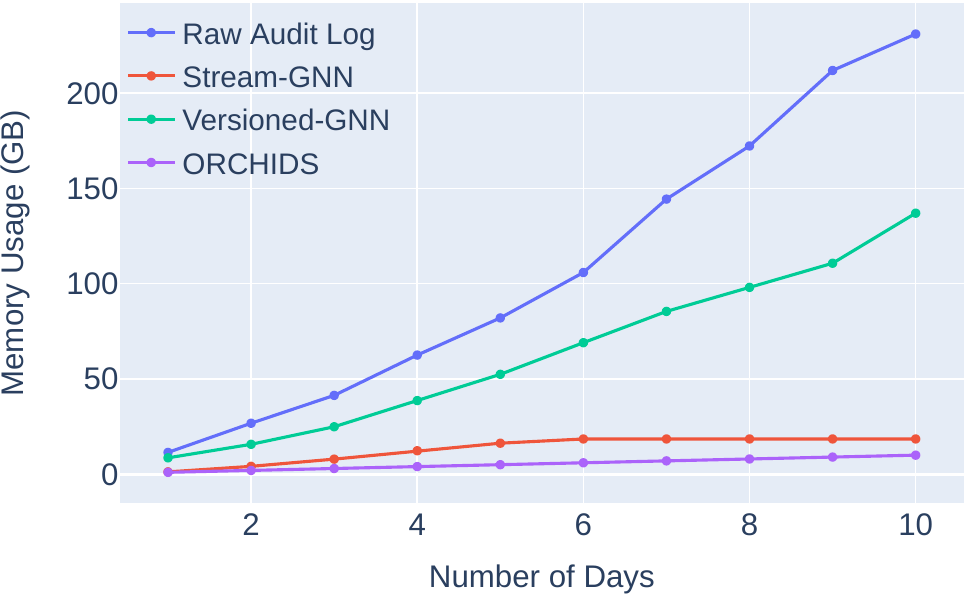}
	\caption{Memory consumption of different IDS models on the Trace dataset, as compared to the raw audit log.
          {\it Versioned-GNN} denotes the memory footprint of Full-GNN if it operated on the more precise versioned
          provenance graph used by \Sys. We were only able to successfully train Full-GNN on $2$ days of
          the versioned graph; subsequent points on this line are estimates.}
        \label{img:memory}
        \vspace{-0.14in}
\end{figure}

%% file: tables/table1.tex
\begin{table}[t]
\centering
\scriptsize
\begin{tabular}{p{3mm}  p{13.5mm}  l*{5}{r} r@{}}
\toprule
\multirow{2}{*}{Task} & \multirow{2}{*}{Model} & \multicolumn{5}{c}{Checkpoint Interval (Num. Edges)} & \multirow{2}{*}{$|V|$} \\
\cmidrule(lr){3-7}&  & {1} & {1K} & {100K} & {1M} & {EOF} & \\
\midrule
\multirow{3}{*}{PP}  & \Sys & - & - & - & -  & 28.9K & 4.08M \\
& Stream-GNN & - & - & - & - & 66.7K & 1.36M \\
& Full-GNN  & -&- & -& -& 66.7K & 1.36M \\
\midrule
\multirow{2}{*}{T} & \Sys & 518K &  &  &  &  &  \\
& Stream-GNN & 62.8K &  &  &  &  &  \\
\midrule
\multirow{2}{*}{ED} & \Sys & 444K & 444K & 444K & 444K & 444K & 4.08M \\
 & Stream-GNN & 160B & 160M & 1.60M & 160K & 400 & 1.36M \\
\midrule
TED & Full-GNN & -& -& -& -& 67.4K & 1.36M \\
\bottomrule
\end{tabular}
\caption{Record the total runtime (in seconds) of Preprocessing (PP), Training (T), Embedding and Detection (ED)
  model tasks using the Trace dataset. TED reflect testing time costs and occur immediately after training for the Full-GNN. Training (T) time refers to time it takes to the constant time it takes to train the model offline. The single number we report for both models at checkpoint interval $1$ is representative for the time it takes to train \Sys or Stream-GNN across all checkpoint intervals.
  Checkpoint Interval reflects that variation in embedding and detection cost based on how many edges are added per test batch,
    with EOF indicating all edges are added at once.
  $|V|$ is the total number of nodes each system encounters in its representation of the underlying graph.}
\label{tbl:computational}
\end{table}

%% file: imgs/figure13.tex
\begin{figure}[t]
  \captionsetup{width=\columnwidth}
  \centering 
	\includegraphics[height=4.5cm]{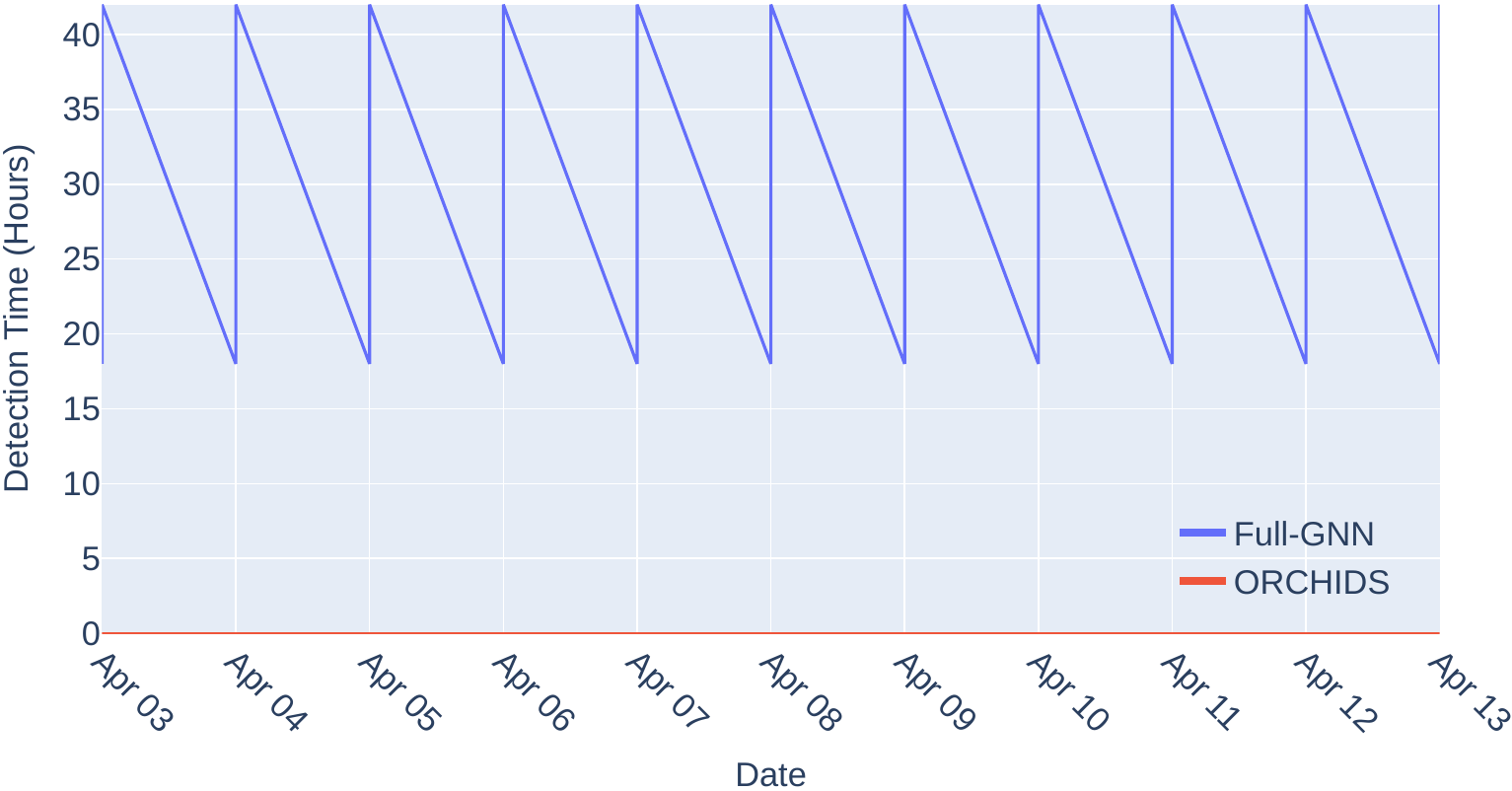}
	\caption{Detection Lag of different IDS models on the Trace dataset.
          Averaged from older runs, {\it Full-GNN} takes $6$ hours to train and analyze audit log. Because of the granularity of the graph, {\it ORCHID} appears to have $0$ lag during detection but {\it ORCHID} requires $0.002$ seconds to process each event.}
	  \label{fig:lag}
\end{figure}

%% file: discussion.tex
\section{Discussion}
ORCHID is a streaming-based Prov-IDS that aims to provide anomaly detection in or close to real-time. However, certain limitations to its design prevent it from being a perfect solution.

\paragraph{Disappearing Memory}
The efficacy of \Sys depends on its ability to represent
  long provenance histories as the system evolves.
Because each embedding is supposed to be a meaningful combination of all the events
  that brought a system entity to its present state  
  the backbone neural network needs to be able to remember that history.
However, previous work has shown that RNNs can forget
  certain paths as they get longer,
  suggesting embeddings that are output by an RNN
  may not reflect all edges that have gone into them.
In practice, this could be weaponized by an attacker that spaces
  out key attack steps to break the causal links within the embedding space. \Sys mitigates this problem by explicitly re-linking root cause entities 
  to their distant successor entities within the embedding.
This mitigation appears to be effective, given the limitations of publicly
  available test data.  
Future work may investigate solutions to boost further RNNs' memory, such that the number of edges forgotten becomes prohibitive.  

\paragraph{Log Retention}
We have touted the ability of \Sys to efficiently and accurately represent
  large provenance graphs for anomaly detection purposes.
However, \Sys' representation of the audit log may not prove as useful
  for downstream tasks like threat hunting and attack forensics.
This means that \Sys is not a replacement for long-term retention of audit
  logs, which are needed to verify and respond
  to the threats detected by \Sys.
At present, the explainability of \Sys classification results still depends
  on an analyst querying the provenance graph of the offending process
  and interpreting the results.
While \Sys facilitates linkability between its detections and the audit log,
  we hope to integrate explainability into the detection model better
   in future work.

%% file: related_works.tex

\paragraph{Streaming Provenance Anomaly Detection}
The issue of processing speed for provenance-based anomaly detection
  has been considered in prior work.
Digest-based vectorization provides an efficient means of quickly representing
  large graphs for cluster analysis.
Manzoor et al.'s Streamspot reports an per-edge throughput of 70$\mu$s
  with a 1000 bit digest \cite{mma2016}.
Han et al.'s Unicorn does not report per-edge throughput, but is shown
  to maintain roughly line speed with the underlying audit framework \cite{hpb+2020}.
However, both Streamspot and Unicorn perform whole-graph/system classification,
  a technique that is known to be vulnerable to adversary evasion \cite{ghb+2023}
  due to the fact that attacks typically account for a miniscule
  proportion of total system activity.
Wang et al.'s ProvDetector performs program classification, but does so
  by analyzing large system subgraphs (e.g. of depth 10) centered on
  the program to be classified \cite{whl+2020}.
Rather than digests, ProvDetector
  down-samples the graph to a small number of paths (e.g., 20)
  and then embeds each using a doc2vec model.
In a 100 endpoint organization,
  the authors estimate that a corpus of 30 programs could be monitored
  in about 5.7 hours, indicating that their system also keeps pace with
  the event stream.
However, by operating on larger subgraphs,
  their system is also more susceptible to evasion \cite{ghb+2023}.\footnote{An additional concern is that multi-path dependencies, e.g., fusions of data, are not possible to express in ProcDetector representations.}
GNN-based approaches like Shadewatcher \cite{zengy2022shadewatcher}
  are computationally efficient, but are only effective if
  training starts after the inference/testing events have already occurred,
  thus making them much slower than their event processing speed would suggest.
All of the above approaches may benefit from online graph learning techniques (e.g., \cite{trivedi2019dyrep})
  as a means of transitioning away from the offline classification model. 
In contrast,
  \Sys can perform real-time processing of the event stream
  while maintaining a lossless representation of the provenance graph
  and monitoring {\it all} programs on the system.

\paragraph{Provenance and Rule-based Intrusion Detection}
Today's Endpoint Detection \& Response (EDR) products rely primarily on rule (heuristic)
 intrusion detection, in which the event stream is pattern matched against
 a set of hand-written queries describing a known attack behavior.
Recent work has observed that pattern matching approaches are also
  applicable for provenance graph representations of the event stream
  \cite{milajerdi2019ieesp, meg+2018, hmw+2017}.
Provenance has also been used to triage/prioritize the alerts
  of traditional endpoint detection products \cite{hassan2019nodoze, hassan2020ieesp},
  demonstrating synergy between event sequence and provenance analysis.
Because these approaches do not require training machine learning models,
  they do not face the same deployment challenges in streaming environments,
  although retaining graphs in memory becomes an issue as they grow \cite{hlj+2020}.

\paragraph{Sequence-based Anomaly Detection}
In contrast to provenance, traditional host anomaly detection
  has modeled the sequences of system calls in an event stream \cite{forrest1996ieeesp,hofmeyr1998intrusion, lee1998usenix, sekar2000ieesp}.
Event sequence analysis continues to enjoy popularity today,
  with anomaly detectors based on neural networks \cite{du2017deeplog,meng2019loganomaly, nedelkoski2020self}.
Because sequence analysis only considers a highly localized window of activity,
  streaming applications are much more straightforward to produce.
However, given that classic sequence-based anomaly detectors were shown to be
  susceptible to mimicry attacks \cite{wagner2002mimicry, tan2002ieesp},
  it is not clear how the introduction of deep learning models has improved their security, if at all.
It may be that analysis of higher-level event streams such as EDR \cite{smv+2018} and NDR \cite{eas+2022} data
  mitigates this problem.

\paragraph{Attack Reconstruction}
Provenance's original applications in security were in {\it attack reconstruction},
  i.e., identifying the steps of an attack once some indicator of compromise
  had already been discovered.
Work in this area includes,
  ATLAS a supervised approach to search for attack sequences in a
  provenance graph using prior knowledge of attacker behaviors \cite{alsaheel2021usenix}.
WATSON clusters provenance subgraphs into higher-level abstract behaviors,
  which can then be used to help analysts
  understand attacker behaviors \cite{zeng2021ndss}.
Similarly, DepComm reduces provenance graphs into communities,
  or smaller subgraphs, that represent some particular behavior \cite{xu2022ieeesp}.
UIScope additionally integrates
  application-layer UI events to understand behaviors \cite{yang2020uiscope}.  
Improvements to intrusion detection algorithms significantly simplify
  attack reconstruction by offering more precise and reliable indicators of compromise,
  as provided by \Sys.
In turn,   
  attack reconstruction systems can aid \Sys by identifying the minority of processes
  in the attack chain that were not initially detected.

%% file: conclusion.tex
\section{Conclusion}
We introduce \Sys, a streaming \Hids capable
  of fine-grained entity-level classification.
\Sys produces node embeddings that
account for all edge relationships in a provenance graph at a fraction of the
memory cost. We show these embeddings achieve competitive anomaly detection
results with state-of-the-art prov-HIDS systems.

%% file: appendix.tex
\appendix
\section{Appendix}
\label{sec:app}

\input{imgs/figure1}

\input{imgs/figure8}

\subsection{Streaming Classification Over Time}


Like any learning system, \Sys must contend with the possibility of concept drift or the shift in distribution between the real world and the data the model was trained on. For host systems, concept drift can be reflected by changes in workloads, where the model of normality used by the IDS to detect abnormal behavior is no long representative of the users behavior. As a result, the abnormality of the user's normal behavior increases leading to higher number of false alerts outputted by the system. Evaluating the effects of concept drift on the performance of the IDS is not easy however, as as modern public intrusion detection datasets only span several days, limiting their ability to be used to exhaustively explore distribution shifts. Moreover, previous work in this area did not examine the effects of concept drift as their training data was representative of underlying distribution that was encountered at test time. To evaluate the effects of concept drift on \Sys, we temporally reduce the size of the training dataset such that it represents only a subset of the behaviors present within the entire dataset; thereby inducing a distribution change between the training and testing set. DARPA Transparent Computing E3 datasets, Theia and Trace represent our longest datasets lasting over 10 days. \Sys is trained on only a single days worth of data, and evaluated over the rest of the dataset.

Figure~\ref{fig:err} reports the mean, and upper-lower quartile for the anomaly scores of benign and attack nodes over the dataset. We report the results of both \Sys and Stream-GNN to compare the effects of concept drift on \Sys against previous work. Like any learning system, Figure~\ref{fig:err} confirms \Sys susceptibility to concept drift represented by the increase in the overall abnormality of the benign data as time progresses. For the Trace dataset (Figs. \ref{fig:eval:err3}-\ref{fig:eval:err4}), the increase in abnormality for benign data stabilizes for both models.\footnote{The spikes in the attack line in these plots mark the beginning of the first attack in the time series.} For the Theia dataset (Fig~\ref{fig:eval:err1}), however, \Sys' benign anomaly score keeps creeping up over the course of the time series. While the mean benign score still falls outside of the attack score's confidence interval, the general trend is concerning. Understanding the disparity between these two datasets requires discussing DARPA Transparent Computing Engagement 3. During the engagement, both performers generated background (benign) activity using a periodically-executing workload script. However, the Theia performers experienced several technical issues that led to machine downtime over the course of the engagement.
Examining the dataset, we believe that the Theia performaers also updated their workload script during each period of downtime. As a result, the background activity in Theia represents three distributions of behavior over the course of the engagement, only one of which appeared in \Sys' preliminary training period. Thus, while Trace heavily advantages the models with its mono-distributional background activity, Theia inadvertently provides a small test of conceptual drift. While it's difficult to say how similar these script changes were to naturalistic drift in machine interactions, given the available data we can at least assert that \Sys continues to effectively classify threats over the course of a 10 day period, at which point retraining may be required.

Consistent with our classification results for this dataset, the confidence intervals of the attack distance scores indicate that \Sys has significantly more discriminative power than Stream-GNN. Moreover the effects of concept drift on Stream-GNN are more pronounced where the upper quartile abnormality score of the benign data surpassed the upper quartile abnormality score of the attack nodes. The difference in robustness to concept drift between Stream-GNN and \Sys can attributed to how each system processes the graph. Stream-GNN considers each node neighborhood in isolation and therefore cannot adapt to great changes to that neighborhood. Meanwhile, \Sys learns to efficiently aggregate incoming information to existing node representations allowing for it to adapt to changes in distribution.

%% file: imgs/figure1.tex
\begin{figure}[t]
  \captionsetup{width=\columnwidth}
  \centering 
	\includegraphics[height=4.5cm]{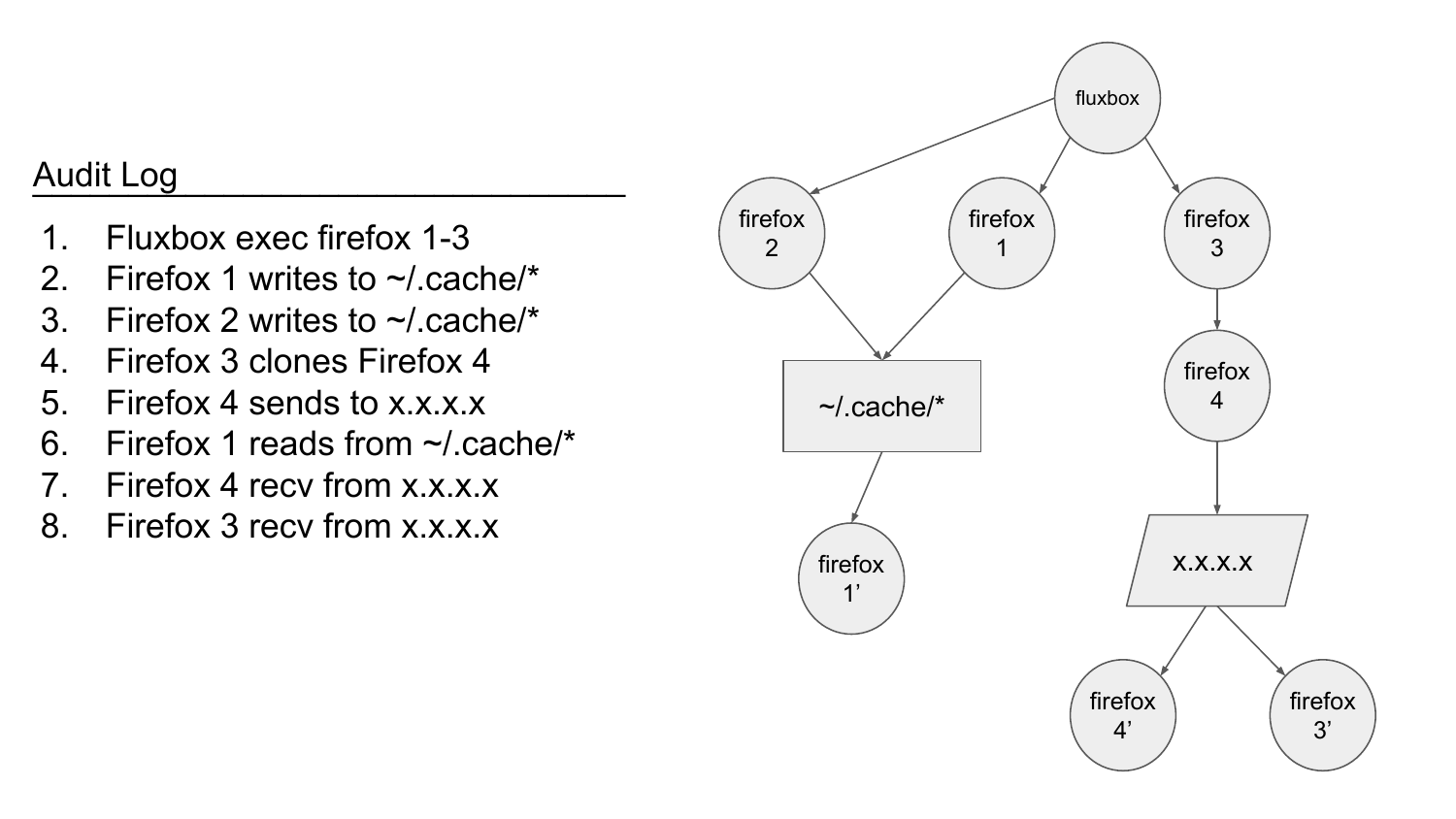}
	\caption{Visualization of a sample audit log and the representative provenance graph from the StreamSpot dataset. The audit log describes an the system interacting with Firefox - specifically the user visiting a website using Firefox. The audit log is ordered by time with each line indicating an additional time step.}
	  \label{img:example:versioning}
\end{figure}

%% file: imgs/figure8.tex
\begin{figure}[t!]
  \begin{subfigure}[]{0.49\linewidth}
    \centering
    \includegraphics[width=1\linewidth]{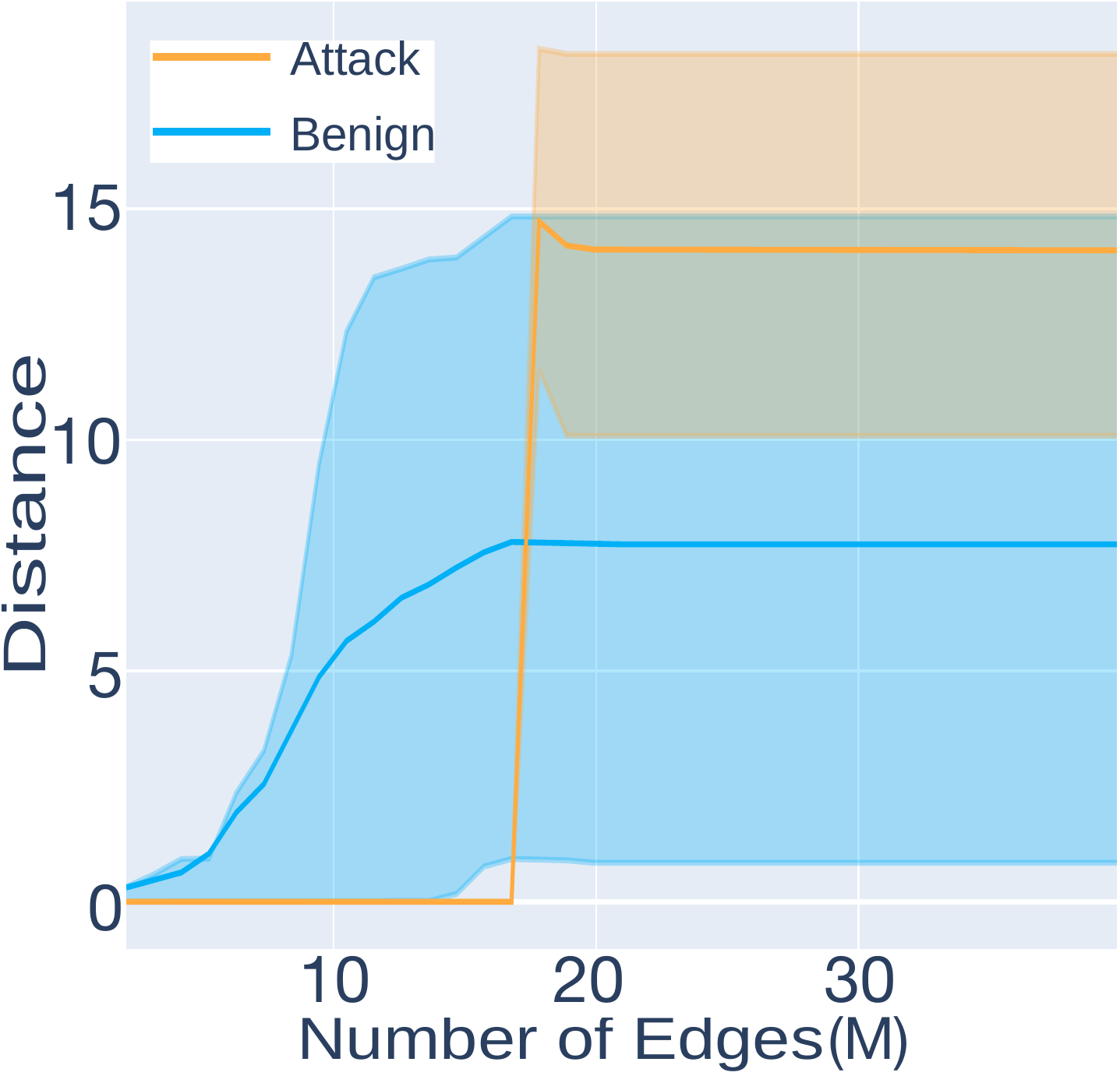}
    \caption{\Sys, Trace}
    \label{fig:eval:err3}
  \end{subfigure}
  \begin{subfigure}[]{0.50\linewidth}
    \centering
    \includegraphics[width=1\textwidth]{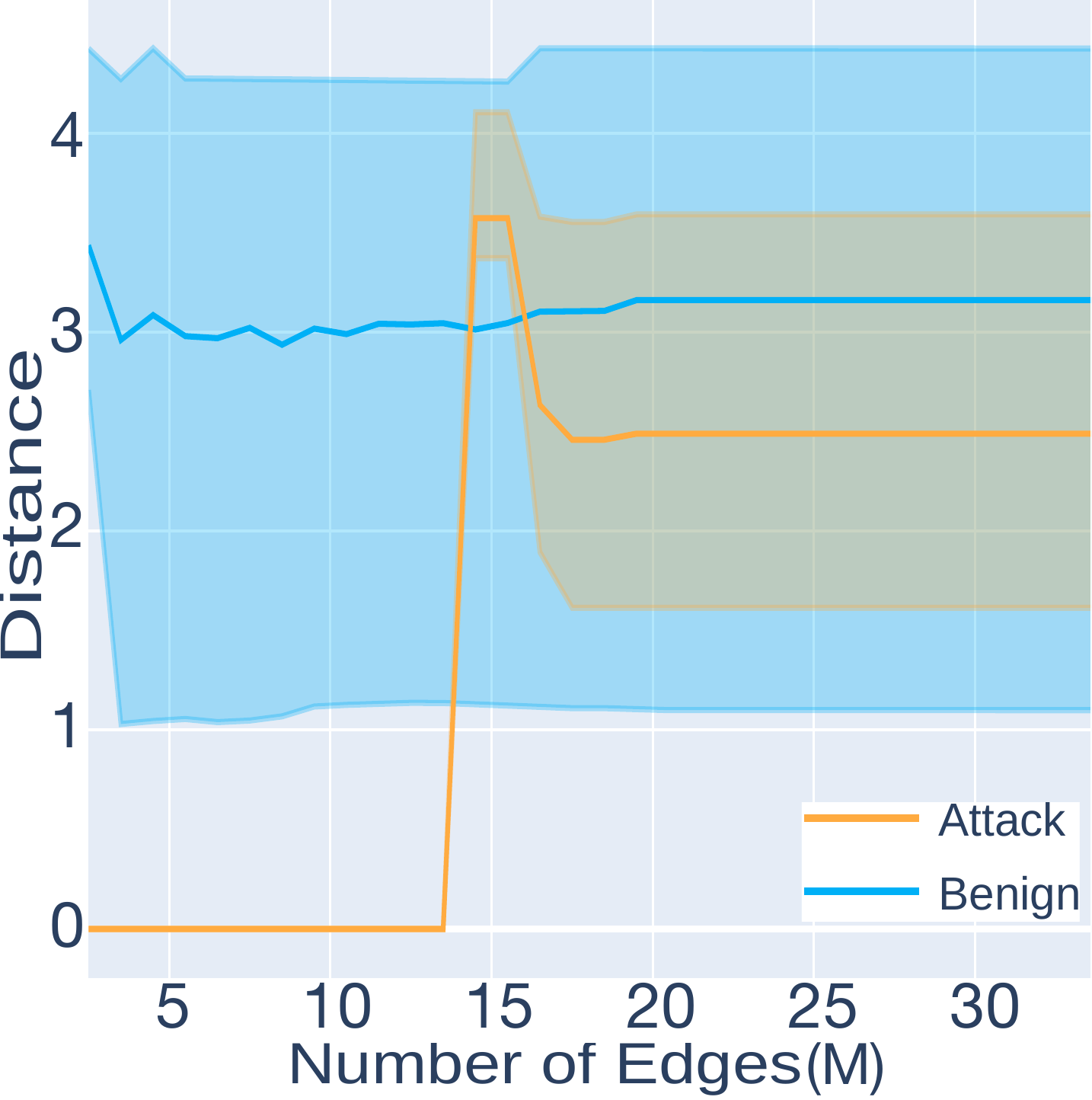}
    \caption{Stream-GNN, Trace}
    \label{fig:eval:err4}
  \end{subfigure}

  \begin{subfigure}[]{0.49\linewidth}
    \centering
    \includegraphics[width=1\linewidth]{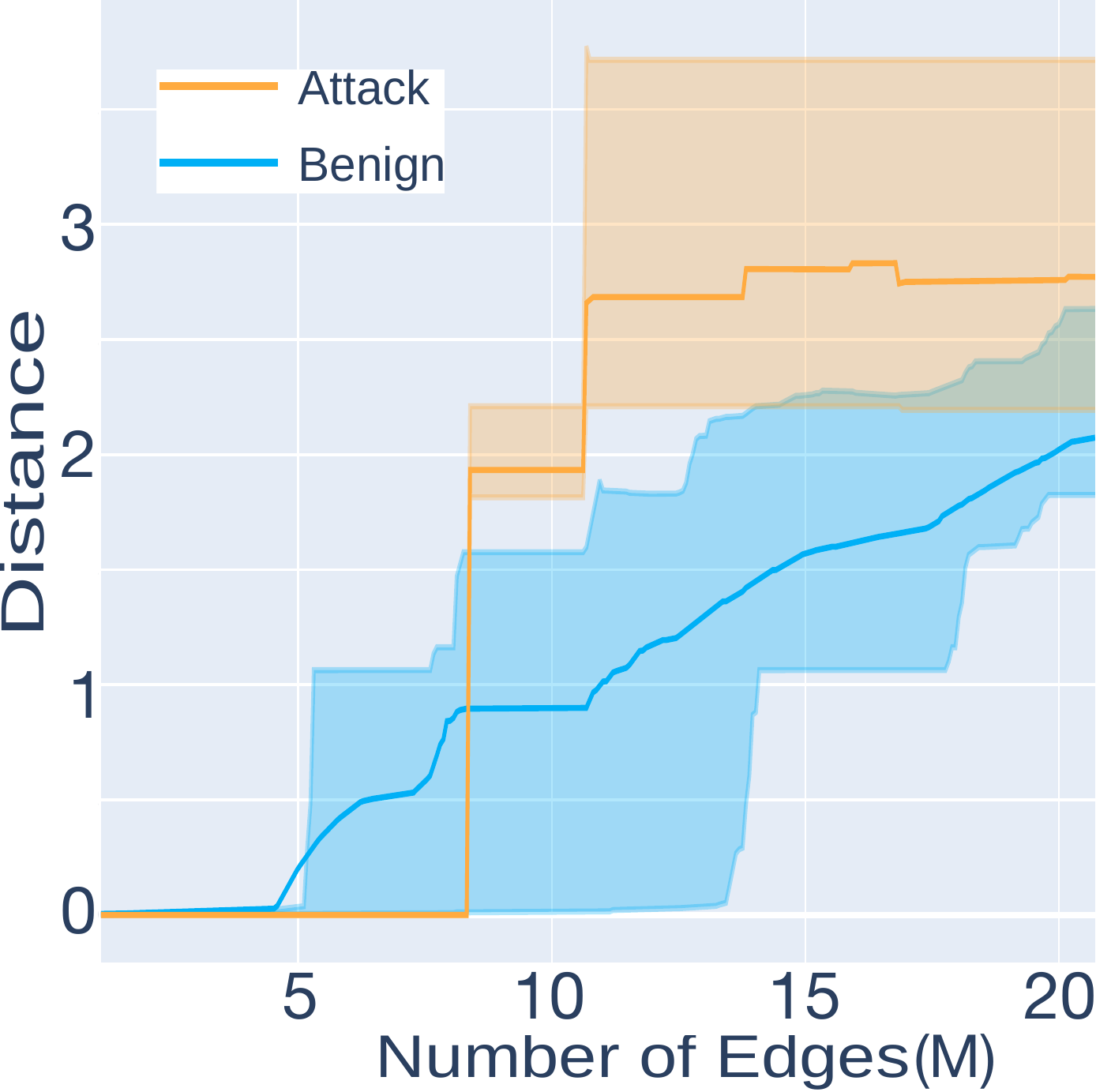}
    \caption{ORCHID, Theia}
    \label{fig:eval:err1}
  \end{subfigure}
  \begin{subfigure}[]{0.50\linewidth}
    \centering
    \includegraphics[width=1\linewidth]{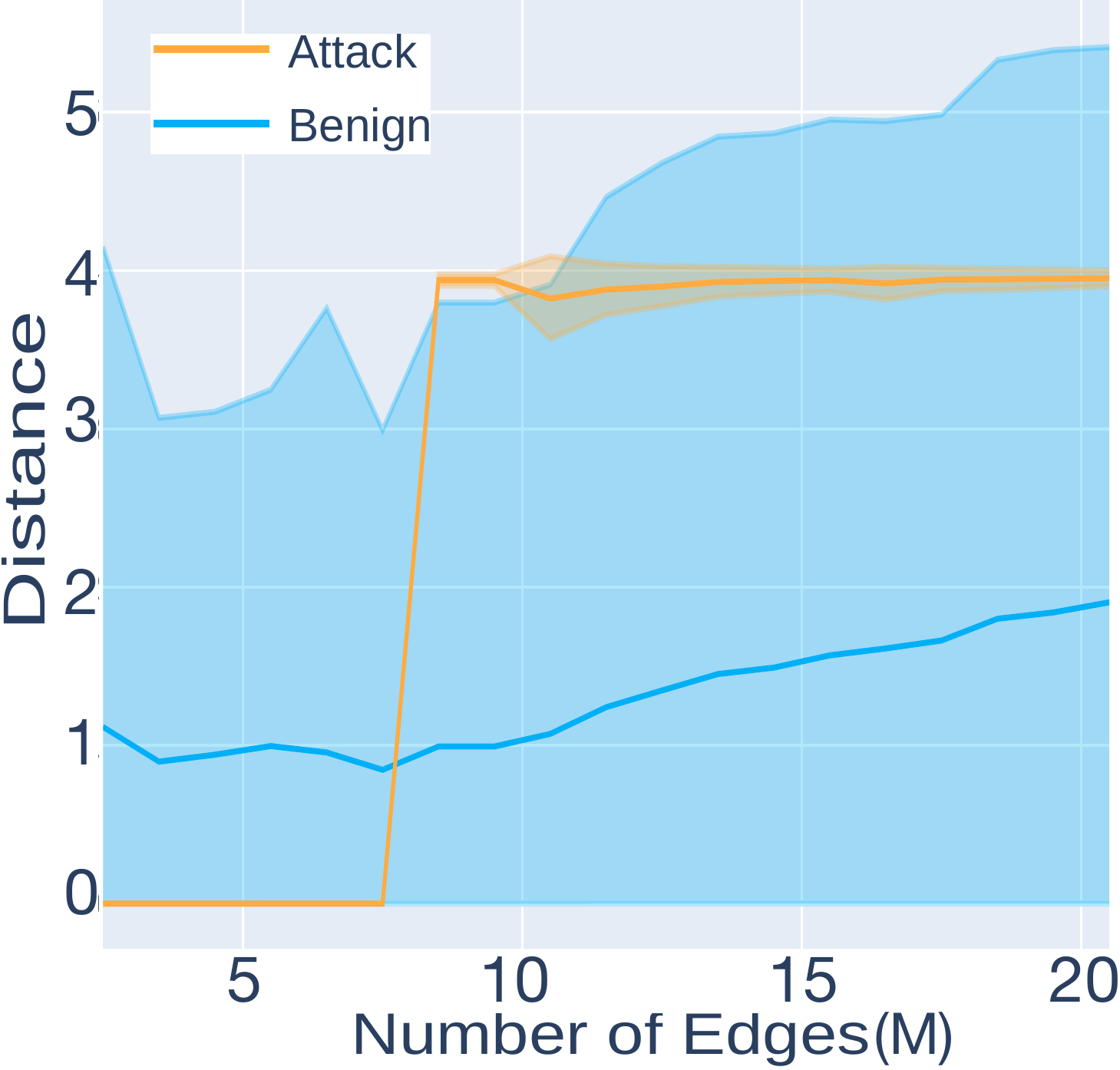}
    \caption{Stream-GNN, Theia}
    \label{fig:eval:err2}
  \end{subfigure}
  \caption{Distribution of nearest neighbor distance between test nodes and benign nodes at increasing time interval. The bottom line is the $25\%$ quartile, the middle line is the mean, and the top line is the $75\%$ quartile.}
    \label{fig:err}
\end{figure}